\documentstyle[eqsecnum,prc,aps]{revtex}
\newcommand{\ovslash}[1]{{\ooalign{\hfil/\hfil\crcr$#1$}}}
\begin{document}
\draft
\title{Neutrino -- nucleon reaction rates in the supernova core \\
in the relativistic random phase approximation}
\author{Shoichi Yamada}
\address{Department of Physics, Graduate School of Science, University 
of Tokyo, \\ 7-3-1, Hongo, Bunkyo-ku, Tokyo 113-0033}
\author{Hiroshi Toki}
\address{Research Center for Nuclear Physics (RCNP), Osaka
University,\\ Ibaraki, Osaka 567-0047, Japan}
\date{\today}
\maketitle
\begin{abstract}
In view of the application to supernova simulations, 
we calculate neutrino reaction rates with nucleons via the neutral and charged
currents in the supernova core in the relativistic 
random phase approximation (RPA) and study their effects on the opacity of 
the supernova core. The formulation is
based on the Lagrangian employed in the calculation of nuclear equation
of state (EOS) in the relativistic mean field theory (RMF). 
The nonlinear meson terms are treated appropriately so
that the consistency of the density correlation derived in RPA with the
thermodynamic derivative obtained from EOS by RMF is satisfied in the
static and long wave length limit. We employ pion and rho meson
exchange interactions together with the phenomenological 
Landau~--~Migdal parameters for the isospin-dependent nuclear
interactions. We find that both 
the charged and neutral current reaction rates are suppressed
from the standard Bruenn's approximate 
formula considerably in the high density regime 
($\rho _{b} \agt 10^{14}$g/cm$^{3}$ with $\rho _{b}$ the baryonic
density). In the low density regime ($\rho_{b} \alt 10^{14}$g/cm$^{3}$), 
on the other hand, the vector current contribution to the
neutrino~--~nucleon scattering rate is enhanced in the vicinity of 
the boundary of the liquid~--~gas phase transition, while the 
other contributions are moderately suppressed there also. In the high 
temperature regime ($T \agt 40$MeV with $T$ the temperature) or in the 
regime where electrons have a large chemical potential, the latter of 
which is important only for the electron capture process and its
inverse process, the recoil of nucleons cannot be neglected and further
reduces the reaction rates with respect to the standard approximate
formula which discards any energy transfer in the processes. 
These issues could have a great impact 
on the neutrino heating mechanism of collapse-driven supernovae.
\end{abstract}
\section{Introduction}
Calculations of neutrino -- nucleon reaction rates in a hot ($T \alt 50$MeV) and
dense ($10^{13}\mbox{g/cm$^{3}$} \alt \rho _{b} 
\alt 10^{15}\mbox{g/cm$^{3}$}$) core of a collapse-driven supernova 
are complicated problems (see, e.g., \cite{ts75,br85,re98a} for 
the standard rates). When the density reaches
$\rho _{b} \sim 10^{13}$g/cm$^{3}$, the average separation of 
nucleons $d \sim 6 \times 10^{-13} \mbox{cm} \ (
\rho_{b} \, / \, 10^{13} \mbox{g/cm$^{3}$}) ^{-1/3} 
$ becomes of the same order as the typical neutrino compton
wave length $\lambda _{\nu} \sim 6 \times 10^{-13}  \mbox{cm} 
\ (E_{\nu} \, / \, 30\mbox{MeV} )^{-1}$. This means that neutrinos are
interacting simultaneously with multiple nucleons for this density or
higher. If nucleons are distributed uniformly in space and
time, which is unlikely, the outgoing waves from multiple targets 
interfere with one another and the cross
sections remain the product of the number of targets and the
cross section of a single scattering. However, in reality, the
distributions of nucleons are fluctuating due to mutual interactions and
the reaction rates will be modified from those obtained with this simple formula. 
Furthermore, the typical energy $E_{\nu} \sim 30 \mbox{MeV}\
(T/10 \mbox{MeV})$  of a neutrino, which is approximately an inverse
of the duration time of interaction between a neutrino and a nucleon, 
is of the same order as the nucleon -- nucleon scattering rate $\Gamma
\sim \langle \sigma v \rangle n_{b} \sim 30 $MeV for the density
$\rho_{b} \sim 3 \times 10^{13}$g/cm$^{3}$ and temperature $T \sim
10$MeV, where $\sigma$, $v$ and $n_{b}$ are the scattering cross
section, the nucleon velocity and the nucleon number
density. This implies that while a neutrino is interacting with a nucleon,
the target nucleon is scattered off by another nucleon. Hence both the
spatial and temporal correlations of nucleons are important.

It is, however, very difficult to calculate these correlations of
nucleons accurately, since they are induced by the strong nuclear
interactions and cannot be treated with a simple perturbation. This
issue was studied by several authors from early on. Sawyer \cite{sa89}, for
example, calculated the density and spin-density fluctuations of nuclear matter 
in the static and long wave length limit from an equation of state, while Iwamoto 
\& Pethick \cite{iw82} investigated them from the Fermi liquid theory. These
results have not been incorporated extensively in simulations of supernovae or
proto neutron stars (see, however, \cite{bl86}) and the simpler
formula mentioned above of the reaction rates
has been used even in the most sophisticated computations.\cite{mz93a,%
mz93b,mz93c,su93}

With increasing recognition that the neutrino transport is one of the
key factors for a successful supernova explosion \cite{jm93,bg93}, this issue has
recently attracted attentions of supernova researchers. In the
neutrino heating scenario of supernova explosion \cite{wi82,bw85}, 
which is supposed to 
be the most promising at present, the shock wave stagnated in the iron 
core after the core bounce is expected to be revigorized by neutrinos
diffusing out of the proto neutron star. It has been shown that it is
important in this mechanism to increase the neutrino heating rate
via the dominant processes of electron neutrino absorption on neutrons 
and electron antineutrino absorption on protons behind the shock wave.
This heating rate is approximately given by:
\begin{equation}
\label{eq:heat}
Q_{\nu} \approx 110 \cdot \frac{L_{\nu,52} \langle E_{\nu,15}^{2}
\rangle}{r_{7}^{2} \langle \mu \rangle} \cdot 
\left \{ 
\begin{array}{c}
Y_{n} \\ 
Y_{p} 
\end{array} 
\right \} 
\quad \left [\frac{\mbox{MeV}}{\mbox{sec} \cdot N_{b}} 
\right ] \quad .
\end{equation}
Here $Y_{n} = n_{n} / n_{b}$ and $Y_{p} = n_{p} / n_{b}$ are the
number fractions of free neutrons and protons, respectively; the 
normalization with the baryonic number density $n_{b}$ indicates 
that the rate per baryon is calculated in Eq.~(\ref{eq:heat});
$L_{\nu,52}$ denotes the neutrino luminosity in units of $10^{52}$erg/s, 
$E_{\nu,15}$ the neutrino energy normalized by $15$MeV, $r_{7}$ the
radius in $10^{7}$cm; $\langle \mu \rangle$ is the mean value of the
cosine of the angle of neutrino propagation relative to the radial
direction. It is clear from this equation that the higher neutrino energy
and/or the greater neutrino luminosity increase the neutrino heating rate.
One obvious way to achieve that is to decrease the neutrino opacities in 
the supernova core. Since the dominant source of opacity for
neutrinos is the neutrino -- nucleon reactions, the modifications of
their rates could have a great impact on the supernova mechanism 
\cite{ke95,ja96,hr98,bs98,bs99,re98b}.

As stated above, the neutrino -- nucleon reaction rates used in the
supernova simulations thus far were mostly evaluated by multiplying the rate for a 
single target nucleon with the nucleon number density (see, e.g.,
\cite{br85,su94} and references therein), thus ignoring 
the correlations due to ambient nucleons. Recently, however, some
authors \cite{bs98,bs99,re98b,re98c,hw91} calculated the
density as well as the spin-density correlations of nucleons due to nuclear
forces based on RPA. As shown later, this method evaluates the
nonuniform nucleon distributions in space and time  
using the mean field approximation. On the other hand,
Raffelt and his company \cite{ke95,ja96,rs95,rs96,rs98}
 insisted that the effect from collisions of two
nucleons on the nucleon spin-density fluctuations cannot be neglected, which is 
not taken into account in RPA (see also \cite{sw89}). 

One of the authors (H. T.) recently published the nuclear EOS based on RMF 
and the Thomas~--~Fermi approximation for finite nuclei covering the wide
range of density, temperature and electron fraction of relevance to
the supernova simulations \cite{sh98,sh99}. In this paper, 
we calculate the nuclear 
correlations in the relativistic RPA based on the Lagrangian used
in the calculation of EOS in RMF \cite{st94a,st94b}. 
It is shown in the next section that this
guarantees the thermodynamical consistency of the neutral vector current part
of reaction rates with EOS in the static and long wave length limit 
if the nonlinear meson terms are appropriately treated. The neutral axial vector
current and the charged current contributions to reaction rates are 
calculated on the same basis but with additional introduction of 
the phenomenological Landau~--~Migdal parameters for the isovector
channel \cite{lm78,hp94,yt98,hp93}, since they do not contribute to RMF 
and are neglected in the
theory. The possible collisional effects which are
supposed to be important particularly in the low density regime and
might have some roles in forming the neutrino spectra will be 
discussed elsewhere \cite{ya99a}, since we consider the consistency of the
reaction rates with EOS which we have currently at hand is more important.

This paper is organized as follows. In the next section we formulate
the neutrino reaction rates with nucleons by the so-called dynamical
structure functions and show that RPA is consistent with EOS obtained
by the mean field theory. Then we represent some modifications of the
reaction rates due to the RPA correlations, using the Bruenn's
standard aprroximation formula as a reference. We
summarize the paper with some discussions of implications for the supernova
simulations in the last section.

\section {Neutrino -- Nucleon Reaction Rates}

First we express the reaction rates in a general form. 
Since we are interested in low energy reactions 
($E \ll M_w$, the mass of weak boson), the weak interaction is well 
approximated by the interaction Lagrangian density:
\begin{equation}
{\cal L}_{I}(x) = \frac{G_{F}}{\sqrt{2}} \  l_{\mu}(x) \  J_{N}^{\mu}(x)
\quad ,
\end{equation}
where $G_{F}$ is the Fermi coupling constant, $l_{\mu}(x)$ is the
lepton weak current given by
\begin{equation}
l_{\mu}(x) = \overline{\psi}_{l}(x) \ \gamma ^{\mu} (1 - \gamma _{5}) 
\ \psi_{l} (x) \quad ,
\end{equation}
and $J_{N}^{\mu}(x)$ is the nucleon counter part, 
\begin{equation}
J_{N}^{\mu}(x) = \overline{\psi}_{N}(x) \ \gamma ^{\mu} (h_{V}
- h_{A} \gamma _{5}) \ \psi_{N} (x) \quad .
\end{equation}
$h_{V}$ and $h_{A}$ are the vector and axial vector coupling constants,
respectively, and are taken for the charged current as $h_{V} = g_{V}
= 1$ and $h_{A} = g_{A} = 1.23$. For the neutral current they are  
$h^{n}_{V} = -1/2$, $h^{n}_{A} = -1/2 \, g_{A}$ and 
$h^{p}_{V} = 1/2 - 2 \sin ^{2} \theta _{W}$, $h^{p}_{A} = 1/2 \, g_{A}$
for neutron and proton, respectively.  $\theta _{W}$ is the
Weinberg angle.  

Following the standard procedure (see, e.g., \cite{ra96}), 
the reaction rates are obtained by
making a square of each matrix element evaluated up to the lowest
order of the Fermi coupling constant, taking the thermal ensemble 
average for the initial state and summing over the final states:
\begin{equation}
\label{eq:rate}
R\,(q^{in},\, q^{out}) \ = \ \frac{G_{F}^{2}}{2} \ 
K_{\alpha \beta}\,(q^{in},\, q^{out}) \ 
S_{N}^{\alpha \beta}(k) \quad .
\end{equation}
$q^{in}$ and $q^{out}$ are the four momenta of the incident
and outgoing leptons, respectively. $k = q^{in} - q^{out}$ is
the four momentum transferred from lepton to nucleon. In the above
equation, the tensor $K_{\alpha \beta}$ comes from the kinematics of 
leptons and given by
\begin{equation}
\label{eq:lepton}
K_{\alpha \beta}\,(q^{in},\, q^{out}) \ = \ 
8 \, (q_{\alpha}^{out} q_{\beta}^{in} \ +\  
q_{\beta}^{out} q_{\alpha}^{in}
\ - \ q^{out} \cdot q^{in} \, g_{\alpha \beta} 
\ + \ i \, \varepsilon _{\alpha \beta}^{\ \ \ \delta \gamma} 
\, q_{\delta}^{out} q_{\gamma}^{in} ) \quad .
\end{equation}  
Here $g_{\alpha \beta}$ is the metric tensor with the signature of 
$[+---]$, and $\varepsilon ^{\alpha \beta \delta \gamma} $ is the 
antisymmetric tensor with $\varepsilon ^{0123} = 1$. All information
of nucleons is contained in the so called dynamical structure function 
$S_{N}^{\alpha \beta}$ defined as:
\begin{equation}
\label{eq:struc}
S_{N}^{\alpha \beta}(k) \ = \ \int \! d^{4} \! x \  e^{ikx} \ 
\langle \, J_{N}^{\alpha}(x) \, J_{N}^{\beta}(0) \, \rangle \quad ,
\end{equation}
where $\langle \, \cdots \, \rangle$ stands for the thermal ensemble average 
of the argument.

The structure function can be generally decomposed as follows due to
the isotropy of the system:
\begin{eqnarray}
S_{N}^{\alpha \beta}(k) \ = \ R_{1}(k) \, u^{\alpha} u^{\beta} \ 
& + & \ R_{2}(k) \, (u^{\alpha} u^{\beta} - g^{\alpha \beta}) \ 
+ \ R_{3}(k) \, k^{\alpha} k^{\beta} \ 
\nonumber \\ \label{eq:gendec}
& + &\ R_{4}(k) \, (k^{\alpha} u^{\beta} + u^{\alpha} k^{\beta}) \ 
+ \ i \, R_{5}(k) \, \varepsilon ^{\alpha \beta \delta \gamma} \,  
u_{\delta} u_{\gamma} \, , 
\end{eqnarray}
where $u^{\alpha }$ is a four velocity of the system. 
Putting Eqs.~(\ref{eq:lepton}) and (\ref{eq:gendec}) into
Eq.~(\ref{eq:rate}), we get 
\begin{eqnarray}
\label{eq:decom}
\nonumber
R\,(E^{in},\, E^{out}, \, \cos \theta) \  = \ 4 \, G_{F}^{2} \,
E^{in} E^{out} \ [ R_{1}(k) \, (1 + \cos \theta) \ 
& + & \ R_{2}(k) \, (3 - \cos \theta) \\ 
& - & \ 2 \, (E^{in} + E^{out}) \, R_{5}(k) \, 
(1 - \cos \theta)] \quad ,
\end{eqnarray}
where $E^{in}$ and $E^{out}$ are the energies of the incident and
outgoing leptons, and $\theta$ is the angle between the incident and outgoing
three momenta. There are three terms with
different angular dependences. In general the third contribution is
much smaller than the other two terms and ignored in the following 
discussions. The meanings of them become
clearer if we take the non-relativistic limit for the nucleon
kinematics. In this limit $R_{1}(k)$ and $R_{2}(k)$ are reduced to
\begin{eqnarray} 
R_{1}(k) & \approx & \ h_{V}^{2} \, \int \! d^{4} \! x \ e^{ikx} \ 
\langle \, \rho_{N}(x) \, \rho_{N}(0) \, \rangle \quad , \\
R_{2}(k) & \approx & \ \frac{h_{A}^{2}}{3} \, \int \! d^{4} \! x \ e^{ikx} \ 
\langle \, \bbox{s}_{N}^{i}(x) \, \bbox{s}_{N}^{i}(0) \, \rangle
\quad . 
\end{eqnarray}
Thus $R_{1}(k)$ comes mainly from the vector currrent part of the nucleon
weak current and is nothing but a density correlation function of 
nucleons. $R_{2}(k)$ originates from the axial vector current of
nucleon and represents a spin-density correlation
function. $\rho_{N}(x)$ is the nucleon density and
$\bbox{s}_{N}^{i}(x)$ is the spin density and the spin components $i$ are
summed up. 

The calculation of the reaction rates is thus reduced to the evaluation of
these correlation functions. It is, however, easier in the field theory
to treat the time ordered product instead of the ordinary one since
we can apply the perturbation theory more easily for the former (see,
e.g., \cite{la87}). Hence 
we define the time ordered product of the weak current 
$\Pi _{N}^{\alpha \beta}(k)$ corresponding
to the dynamical structure function as:
\begin{equation}
\label{eq:defcau}
i \, \Pi _{N}^{\alpha \beta}(k) \ = \ \int \! d^{4} \! x \  e^{ikx} \ 
\langle \, T \, J_{N}^{\alpha}(x) \, J_{N}^{\beta}(0) \, \rangle \quad ,
\end{equation}
where the symbol $T$ stands for the chronological ordering of
operators. It is  also convenient to consider the corresponding
retarded Green function defined as 
\begin{equation}
\label{eq:defret}
i \, \Pi _{N}^{R \alpha \beta}(k) \ = \ \int \! d^{4} \! x \  e^{ikx} \ 
\Theta (t) \ 
\langle \, [ J_{N}^{\alpha}(x), \, J_{N}^{\beta}(0)] \, \rangle \quad ,
\end{equation}
where $[A, B]$ denotes the commutator of $A$ and $B$ and $\Theta (t)$ is the
Heaviside function. In fact, the dynamical structure function is related with the
imaginary parts of those Green functions via a simple formula: 
\begin{equation}
\label{eq:disp}
S_{N}^{\alpha \beta}(k) \ = \ - \, \frac{2}
{1 + e^{- \beta \, (k_{0} - \Delta \, \mu})} \ Im \, \Pi_{N}^{\alpha \beta}
\ = \ - \, \frac{2}
{1 - e^{- \beta \, (k_{0} - \Delta \, \mu})} \ Im \, \Pi_{N}^{R \alpha \beta}
\quad .
\end{equation}
Here $\Delta \mu = \mu _{out} - \mu _{in}$ is the difference of the 
chemical potentials between the outgoing and incident
nucleons. Thus what we have to do is to somehow calculate 
Eqs.~(\ref{eq:defcau}) or (\ref{eq:defret}).

\subsection{Thermodynamic Consistency}

In this section, we discuss the thermodynamic consistency of the
approximation for reaction rates with that for
EOS. By consistency we mean the following: The reaction rates are
nothing but correlation functions as stated above. For example, the
density correlation function reduces in the static and long wave length
to $\langle (N - \langle N \rangle ) \, 
\cdot \, (N - \langle N \rangle ) \rangle \, / \, V^{2} = ( \langle N^{2} 
\rangle - \langle N \rangle ^{2}) \, / \, V^{2}$. Here $N$ and $V$ are the
baryonic number and the volume of the system, respectively. This
thermal ensemble average is related to the
thermodynamic derivative of the number density with
respect to the chemical potential as $\left ( \partial N / 
\partial \mu \right ) _{T} = \beta \, \langle (N - \langle N \rangle ) \, 
\cdot \, (N - \langle N \rangle )$ with $\beta = 1 / T$, which is
obtained from EOS \cite{kb62}. This sort of 
relation should be satisfied also in approximate formulations, since it 
guarantees the correct behavior of reaction rates in this limit. 
In the following we show that RPA is consistent in 
this sense with the mean field theory \cite{bs98}. The argument is mainly 
indebted to the papers \cite{bk61,by62}. For simplicity we consider 
the non-relativistic density correlation in 
the imaginary time formalism. The extension to the relativistic case
\cite{vb98} and/or to the real time formalism is possible \cite{iv98}.

EOS can be calculated once the number density is obtained for a given
temperature and chemical potential. The number density in turn is
obtained from the single particle Green function as 
$n_{b} (x_{1}) = - i \, G(1, 1^{+})$. Here
the Green function is defined as usual, $i \, G(1, 2) = \langle T \, \phi (x_{1})
\phi ^{\dagger} (x_{2}) \rangle$, and $1^{+}$ in the argument denotes the limit
$t_{2} \rightarrow t_{1} + 0$. Hence the approximation for EOS can be
regarded as the approximation of the Green function, or that of the
self energy $\Sigma$ since the Green function is determined by the
Dyson equation $G^{-1} = G_{0}^{-1} - \Sigma$, where $G_{0}$ is the
free propagator. 

The response of the system to the external disturbance is studied by adding the
extra interaction like, \\ ${\cal L}_{ext} = 
\int ^{-i \beta} _{0} \! d\bbox{x_{1}} dt_{1} d\bbox{x_{2}} dt_{2}
\, \phi ^{\dagger} (\bbox{x_{2}}, t_{2}) 
\, U(\bbox{x_{2}}, t_{2}, \bbox{x_{1}}, t_{1}) 
\, \phi (\bbox{x_{1}}, t_{1})$, where only the time integration region
is shown explicitly. Then the response function is given by
\begin{equation}
\label{eq:defl}
\left . \frac{\delta G(1, 1';U)}{\delta U(2', 2)} \right |
_{U = 0} \ = \ - \ \left [ G_{2}(1, 1', 2, 2') - G(1, 1') \, 
G(2, 2') \right ] \ = \ i \, L(1, 1', 2, 2') \quad , 
\end{equation}
where $G_{2}(1, 1', 2, 2')$ is the two particle Green function, and 
$G(1, 2; U)$ is the single particle Green function under the
external disturbance and is written in the interaction representation
as 
\begin{equation}
\label{eq:defgu}
i \, G(1, 2; U) \ = \ \frac{\langle T \ S \, \phi (x_{1}) \, 
\phi ^{\dagger}(x_{2}) \rangle}
{\langle T \ S \rangle} \quad 
\end{equation}
with $S = \exp (- i {\cal L} _{ext})$.  
Eq.~(\ref{eq:defl}) is derived directly by taking a derivative of 
Eq.~(\ref{eq:defgu}). The retarded Green function defined by 
Eq.~(\ref{eq:defret}) is obtained from the response function 
$L(1, 1^{+}, 2, 2^{+})$ by
the analytic continuation \cite{kb62}. The equation for the response function 
$L(1, 1', 2, 2')$ is obtained by taking a functional derivative 
of the Dyson equation as
\begin{equation}
\label{eq:leq}
L(1, 1', 2, 2') \ = \  L_{0}(1, 1', 2, 2') \ - \ 
\int \! \! dx_{3} dx_{3'} dx_{4} dx_{4'} \ 
L_{0}(1, 1', 3, 3') \, \cdot \, \Xi (3', 3, 4', 4) \, \cdot \, L(4, 4', 2, 2') 
\end{equation}
with 
\begin{eqnarray}
i \, L_{0}(1, 1', 2, 2') \ & = & \ G(1, 2') \, G(2, 1') \\
\label{eq:defxi}
i\, \Xi(1', 1, 2', 2) \ & = & \ \frac{\delta \Sigma (1', 1)}{\delta G(2, 2')}
\quad .
\end{eqnarray}
It was shown \cite{by62} that if the self-energy $\Sigma$ is 
derived from a potential as 
$\Sigma (1', 1) = \delta \Phi / \delta G(1, 1')$, or the
integrability condition $\delta \Sigma (1', 1) / \delta G(2, 2') = 
\delta \Sigma (2', 2) / \delta G(1, 1')$ is satisfied, then the approximation 
is conservative, that is, the approximate number density as well as
the response function so obtained
satisfies the conservation equation just like the exact one. 
This result in turn dictates the behavior
of the response function in the static and long wave length limit as
follows. The response function is in general given by 
\begin{equation}
\label{eq:lsp}
L(\Omega, \bbox{k}) \ = \ \frac{1}{2 \pi } \int _{-\infty} ^{\infty}
\! d k_{0} \ 
\frac{L^{>}(k_{0}, \bbox{k}) - L^{<}(k_{0}, \bbox{k})}
{\Omega - k_{0}} \ = \ \frac{1}{2 \pi } \int _{-\infty} ^{\infty}
\! d k_{0} \
\frac{(1 - e ^{-\beta k_{0}}) \, L^{>}(k_{0}, \bbox{k})}
{\Omega - k_{0}} \quad .
\end{equation}  
Here $L(\Omega, \bbox{k})$ is a Fourier transform of $L(1, 1^{+}, 2,
2^{+})$ with respect to $x_{1} - x_{2}$ and $\Omega$ is an imaginary 
energy. $L^{>}(k_{0}, \bbox{k})$ and $L^{<}(k_{0}, \bbox{k})$ are
analytic functions with respect to $k_{0}$  and are Fourier
transformed from $L^{>}(x_{1}, x_{2}) = \langle (n_{b}(x_{1}) -
\langle n_{b}(x_{1}) \rangle ) \, \cdot \, (n_{b}(x_{2}) - 
\langle n_{b}(x_{2}) \rangle ) \rangle$ and $L^{<}(x_{1}, x_{2}) 
= \langle (n_{b}(x_{2}) - \langle n_{b}(x_{2}) \rangle ) \, \cdot \,
(n_{b}(x_{1}) - \langle n_{b}(x_{1}) \rangle ) \rangle$, respectively.
Thus $L^{>}(x_{1}, x_{2})$ is nothing but a density correlation function we
are seeking for. In fact, we use this equation (\ref{eq:lsp}) to
derive the density correlation function $L(\Omega, \bbox{k})$. 
If the approximate response function satisfies the conservation law, 
$L^{>}(k_{0}, 0) = 2 \pi \, \delta( k_{0} ) \, \tilde{L}^{>}(0,0)$ 
in the long wave length
limit, $|\bbox{k}| \rightarrow 0$. This then results in the equation
\begin{equation}
\label{eq:llong}
\lim_{|\bbox{k}| \rightarrow 0} L(0, \bbox{k}) \ = \ 
- \, \beta \, \tilde{L}^{>}(0, 0) \ = \ - \, \beta \, 
V \, \langle (n_{b} - 
\langle n_{b} \rangle ) \, \cdot \, (n_{b} - 
\langle n_{b} \rangle ) \rangle \quad .
\end{equation}

On the other hand, the thermodynamical derivative, $\left (\partial n_{b} / 
\partial \mu \right )_{T} = - i \, \partial G(1, 1^{+}) / \partial \mu$, 
is evaluated from the Dyson equation as
\begin{eqnarray}
\nonumber
\frac{\partial G(1, 1')}{\partial \mu} \ & = & \ 
- \ \int \! \! dx_{2} dx_{2'} \ G(1, 2') \, \cdot \, \frac{\partial G^{-1}(2', 2)}
{\partial \mu } \, \cdot \, G(2, 1') \\
\nonumber
& = & \ - \ \int \! \! dx_{2} dx_{2'} \ G(1, 2') \ 
\left [ \delta(2' - 2) - \frac{\partial \Sigma (2', 2)}{\partial \mu}
\right ] \ G(2, 1') \\
& = & \ - \ \int \! \! dx_{2'} \ G(1, 2') \, \cdot \, G(2', 1') \
\nonumber \\ 
& & \ + \ \int \! \! dx_{2} dx_{2'} dx_{3} dx_{3'} \ 
G(1, 2') \, \cdot \, i \, \Xi (2', 2, 3', 3) \, \cdot \, 
G(2, 1') \, \cdot \, \frac{\partial G(3, 3')}{\partial \mu} .
\end{eqnarray} 
Solving this equation using Eqs.~(\ref{eq:leq}),~(\ref{eq:llong}), we get
\begin{equation}
\left (\frac{\partial N}{\partial \mu } \right )_{T} \ = \ - \, i \, V \, 
\frac{\partial G(1, 1^{+})}{\partial \mu } \ = \ V \, 
\lim_{|\bbox{k}| \rightarrow 0} L(0, \bbox{k}) \ = \ 
\beta \, \langle (N - \langle N \rangle ) (N - \langle N \rangle ) \rangle
\quad .
\end{equation}

This is the equation which represents the thermodynamical
consistency between the response function and EOS. So far the only
assumption is that we have used the same approximate self energy  
for EOS and the response function. Hence the above
argument can be applied to any conserved current and conserving
approximation. In the next
section, we consider a specific approximation of the self energy which 
leads to the mean field theory and RPA.

\subsection{RPA}

We base our discussion on the Lagrangian of the relativistic mean
field (RMF) theory:
\begin{eqnarray}
{\cal L}_{RMF} & \  = \ & \overline{\psi}_{N} \, (\,
i \, \gamma_{\mu} \, \partial ^{\mu} \, - \, M_{N} \, - \, 
g_{\sigma} \, \sigma 
\, - \, g_{\omega} \, \gamma_{\mu} \, \omega ^{\mu}
\, - \, g_{\rho} \, \gamma_{\mu} \, \tau^{a} \, \rho^{a \mu} \, ) \,
\psi_{N} \nonumber \\
& \ + \ & \frac{1}{2} \, \partial _{\mu} \, \sigma \, \partial ^{\mu} \, \sigma
\, - \, \frac{1}{2} \, m_{\sigma}^{2} \, \sigma ^{2}
\, - \, \frac{1}{3} \, g_{2} \, \sigma ^{3} \, - \, \frac{1}{4} \,
g_{3} \, \sigma ^{4} \nonumber \\
& \ - \ & \frac{1}{4}\, F_{\mu\nu} \, F^{\mu\nu} \, + \, 
\frac{1}{2}\,  m_{\omega}^{2}
\, \omega _{\mu} \, \omega ^{\mu} \, + \, \frac{1}{4} \,  c_{3} \,
(\omega _{\mu} \, \omega ^{\mu}) ^{2} \nonumber \\
& \ - \ & \frac{1}{4} \, G_{\mu\nu}^{a} \, G^{a\mu\nu} \, + \,
\frac{1}{2} \, m_{\rho}^{2} \, \rho _{\mu} ^{a} \, \rho ^{a\mu}
\quad ,
\end{eqnarray}
where the standard notations are used: $\psi _{N}$ denotes a baryonic
field of mass $M_{N}$. $\sigma$, $\omega _{\mu}$ and 
$\rho ^{a} _{\mu}$ are $\sigma $, $\omega $ and $\rho $ meson fields
with $m_{\sigma } = 511$MeV, $m_{\omega } = 783$MeV and 
$m_{\rho } = 770$MeV, respectively. 
$F_{\mu\nu}$ and $G_{\mu\nu}^{a}$ are the antisymmetric field tensors
for $\omega $ and $\rho $ mesons. The constants $g_{\sigma} = 10.0$,
$g_{\omega } = 12.6$ and $g_{\rho } = 4.6$ are the coupling constants for
the interactions between mesons and nucleons. $g_{2} = -7.2 
\mbox{fm$^{-1}$}$, $g_{3} = 0.62$ and 
$c_{3} = 71.3$ are the self-coupling constants for $\sigma $ and $\omega $
mesons. The quoted numbers are taken from the TM1 parameter set
published by Sugahara \& Toki \cite{st94b}(see also \cite{st94a}).
 
The approximate self-energy employed in the mean field theory is given
as follows. For simplicity, only the contribution from $\sigma $
meson, that is the scalar part, is shown. The extension to other meson 
contributions is done in an analogous way.
\begin{equation}
\label{eq:aprx}
\Sigma _{s} (1', 1) \ = \ g_{\sigma } \, \delta (1 - 1') \, \cdot \, 
\sigma (1) \quad ,
\end{equation}
where $\sigma$ meson is assumed to be a classical field and obeys the
equation of motion:
\begin{equation}
\label{eq:sig}
(\partial _{\nu } \partial ^{\nu } \ + \ m_{\sigma }^{2}) \, \sigma
\ + \ g_{2} \, \sigma ^{2} \ + \ g_{3} \, \sigma ^{3}
\ = \ - \ \Delta _{0} ^{-1} \, \sigma  
\ + \ g_{2} \, \sigma ^{2} \ + \ g_{3} \, \sigma ^{3}
\ = \ - \, g_{\sigma } \langle \, \overline{\psi } _{N} \psi _{N} \, \rangle
\quad .
\end{equation} 
Here $\Delta _{0} ^{-1}$ is the inverse of the free
propagator. Solving Eqs.~(\ref{eq:aprx}), (\ref{eq:sig}) with the
Dyson equation of nucleon consistently for stationary and uniform matter, we
obtain RMF for the nuclear EOS. 

On the other hand, using the same self-energy 
and Eq.~(\ref{eq:sig}), we can evaluate $\Xi$ defined by Eq.~(\ref{eq:defxi}) as 
\begin{equation}
\Xi(1', 1, 2', 2) \ = \ g_{\sigma }^{2} \, \delta (1' - 1) \, 
\delta (2' - 2) \, \cdot \, \left [ - \, \Delta _{0}^{-1}(1 - 2)  \, + \, 
2 \, g_{2} \, \overline{\sigma } \, + \, 3 \, g_{3} \, \overline{\sigma} ^{2}
\right ]^{-1} \, \delta (1 - 2) \quad ,
\end{equation}
where $\overline{\sigma}$ is the expectation value of $\sigma $ in the static
uniform matter. It is obvious that this approximation is conserving in 
the above sense. Putting this into Eq.~(\ref{eq:leq}), we get
the equation for $L(1, 1^{+}, 2, 2^{+})$ as
\begin{equation}
\label{eq:eqsigma}
L(1, 1^{+}, 2, 2^{+}) \ = \ L_{0}(1, 1^{+}, 2, 2^{+}) \ - \ 
\int \! \! dx_{3} dx_{4} \ L_{0}(1, 1^{+}, 3, 3^{+}) \, \cdot \, 
\frac{- \, g_{\sigma }^{2}}{\tilde{\Delta } ^{-1} (3 - 4)} \, \cdot \,
L(4, 4^{+}, 2, 2^{+}) \quad .
\end{equation}
Here $\tilde{\Delta }^{-1} \, = \, \Delta _{0}^{-1} 
\, - \, 2 \, g_{2} \, \overline{\sigma } \, - \, 
3 \, g_{3} \, \overline{\sigma} ^{2}$. This is well known RPA except
for the modification of the meson propagator due to its
self-coupling. 
Thus we established the thermodynamic consistency between RMF and RPA.

It should be noted that the modified meson propagator is also obtained by
expanding the Lagrangian around the stationary point up to the
quadratic order of perturbation \cite{hn89,ma97a,ma97b,ma97c}. 
Since the stationary solution is nothing 
but the mean field, the correlation is described by the harmonic
oscillation around the mean field in the above approximation. In this
sense, it is natural that we get RPA consistent with RMF in the static 
and long wave length limit.

The above argument is in a rigorous sense applicable only to 
the conserved quantity such
as the baryonic number. In general the axial vector part of the
nucleon weak current is not a conserved current. However, we extend
the above method to the axial vector contribution, since we can
thus argue both vector and axial vector contributions on the same
basis. Moreover, the interpretation that we describe the correlations by the 
harmonic oscillations around the static and uniform mean field is
still valid. Thus the basic equations of RPA are summarized as
\begin{eqnarray}
\label{eq:rpaeq}
Tr \! \left [ \, \Gamma ^{a} \, L(1 - 2) \, \Gamma ^{b} \, \right ]
& = & \ Tr \! \left [ \, \Gamma ^{a} \, L_{0}(1 - 2) \, \Gamma ^{b} \, \right ] 
\nonumber \\
& - & \ \int \! \! dx_{3} dx_{4} \sum _{\Gamma ^{c}} \,
Tr \! \left [ \, \Gamma ^{a} \, L_{0}(1 - 3) \, \Gamma ^{c} \, \right ]
\, \cdot \, V^{c}_{pot}(3 - 4) \, \cdot \, Tr \! \left [ \, 
\Gamma ^{c} \, L(4 - 2) \, \Gamma ^{b} \, \right ] 
\quad .
\end{eqnarray} 
Here $L(1 - 2)$ and $L_{0}(1 - 2)$ are the abbreviations of 
$L(1, 1^{+}, 2, 2^{+})$ and  $L_{0}(1, 1^{+}, 2, 2^{+})$,
respectively. $\Gamma $'s denote gamma matrices, $\bbox{1}, 
\gamma ^{\mu}, \gamma ^{\mu} \gamma _{5}$, and the matrix structure of 
the response functions is explicitly indicated by the trace
operations. $V_{pot}$ is the nuclear potential mediated by mesons and
is given by the modified propagators as
\begin{equation}
V_{pot} ^{s} (1 - 2) \ = \ - \, g_{\sigma }^{2} \, \tilde{\Delta } (1 - 2)
\end{equation}
for $\sigma$ meson, for example. The response function $L(1 - 2)$ is
analytically continued to the retarded Green function, which in turn
gives the structure functions by Eq.~(\ref{eq:disp}). 
It is also possible to work in the real
time formalism from the beginning, which we did in this
paper. Regardless, they are equivalent to each other \cite{gu94}. The explicit
expressions of $Tr \! \left [ \, \Gamma ^{a} \, L_{0}(1 - 3) \, \Gamma ^{c} \,
\right]$ are given in Appendix.

It is clear from the above derivation that RPA is an approximation 
obtained from the self-energy evaluated up to the first order of the coupling 
constant and that processes such as collisions of nucleons are ignored.
This can be understood from the fact that RPA is also obtained from the
collisionless Boltzmann equation \cite{kb62}. The collisional effect might be
important particularly in the low density regime \cite{rs95,rs96,sw89} 
and could in principle be included in the present formulation 
by taking  higher order corrections of the self-energy \cite{kr86}. 
It is, however, very difficult to do this in practice. In this
paper we make more of the
consistency of the approximations used for reaction rates and EOS 
we have currently at hand. The study of possible effects of
scatterings which was discussed in the introduction will be
published elsewhere \cite{ya99a}. 

In RMF we ignore the negative energy contribution to the nucleon
spinor. In so doing, the positive energy part is defined in a
density-dependent way. This extra density dependence should have
made Eqs.~(\ref{eq:eqsigma}), (\ref{eq:rpaeq}) more
complicated, which we ignored in this paper. It should be noted that this
inconsistency is quite minor except for small parametr regions very
close to the phase boundary with which we are not concerned here.
  
\subsection{Residual Interactions} 

The Lagrangian shown above does not include the contributions of the pion 
and the tensor coupling of the $\rho$ meson since they do not contribute
to the mean field \cite{st94a,st94b}. To describe 
fluctuations in spin and isospin around the mean field 
these residual interactions should be added to the above Lagrangian as \cite{bm90}
\begin{equation}
{\cal L}_{res} \ = \ {\cal L}_{\pi } \, + \, {\cal L}_{\rho } \ = \ 
- \  \overline{\psi } _{N} \, \left [ \, i \, \frac{g_{\pi }}{2 M_{N}} \, 
\gamma ^{\mu } \, \gamma _{5} \, \partial _{\mu} \pi ^{a}
\ - \ \frac{g_{\rho } \, C_{\rho }}{2 M_{N}} \, \sigma _{\mu \nu } \,
\partial ^{\nu} \rho ^{a \mu } \, \right ] \tau ^{a} \,
\psi _{N} \quad 
\end{equation}
with $g_{\pi }^{2} / 4 \pi = 14.08$, $g_{\rho }^{2} / 4 \pi = 0.41$
and $C_{\rho } = 6.1$. Accordingly $\sigma ^{\mu \nu}$ is added to
$\Gamma $'s in Eq.~(\ref{eq:rpaeq}). 

It is also known that there are short ranged repulsive correlations   
for these isovector channels. They are conveniently included 
by the phenomenological Landau~--~Migdal parameters \cite{lm78}. 
In this paper they are
implemented, following Horowitz \& Piekarewicz \cite{hp94,hp93}, 
by modifying the meson propagators as
\begin{eqnarray}
\frac{1}{q_{\mu }^{2} \, - \, m_{\pi }^{2}} & \  \rightarrow \ & 
\left [ \, \frac{1}{q_{\mu }^{2} \, - \, m_{\pi }^{2}} 
\ - \ \frac{g'_{\pi }}{q_{\mu }^{2}} \, \right ] \\
\frac{1}{q_{\mu }^{2} \, - \, m_{\rho }^{2}} & \  \rightarrow \ & 
\left [ \frac{1}{q_{\mu }^{2} \, - \, m_{\rho }^{2}} 
\ - \ \frac{g'_{\rho }}{q_{\mu }^{2}} \, \right ]
\end{eqnarray}
with the Landau~--~Migdal parameters $g'_{\pi } = 0.70$ and $g'_{\rho } = 0.30$.
Unlike the conventional non-relativistic treatment, we take different
parameter value for $\rho $ mesons from pions, which better reproduces the
electron scattering data as well as the spin-transfer observables in 
$(p, n)$ reactions \cite{hp94,yt98,ba83,ch93,sa94,hp93}.

As for the electromagnetic interaction, we make the Thomas-Fermi
approximation, where the photon propagator is replaced by the
screened one as
\begin{equation}
\frac{g_{\mu \nu }}{- \, q_{\mu }^{2}} \qquad \rightarrow \qquad 
\frac{g_{\mu \nu }}{q_{TF}^{2} \, - \, q_{\mu }^{2}}
\end{equation}
with $q_{TF}^{2} = 4 \, e^{2} \, \pi ^{1/3} (3 \, n_{e})^{2/3}$, where $e$ 
and $n_{e}$ are the electron charge and number density, respectively.
Then it was added to the $V_{pot}$'s in Eq.~(\ref{eq:rpaeq}).
  
It is finally noted that since 
the neutrino couples only to the spin-transverse correlations in RPA, the pion
(spin-longitudinal correlations) does not contribute to the neutrino 
reaction rates considered here.

\section{Results}

\subsection{Neutral Current Reactions}

In this section, we discuss  correlation effects on the neutral current 
reactions, that is, neutrino -- nucleon scatterings. The high density regime, 
$\rho _{b} \geq 2 \times 10^{14}$g/cm$^{3}$, and the low density regime, 
$\rho _{b} < 2 \times 10^{14}$g/cm$^{3}$, are discussed separately, 
since the behaviors of $R_{1}(k)$ in Eq.~(\ref{eq:decom}), which is
refered to as  the vector current correlations in the following, 
are qualitatively different between the two regimes. 

The vector current correlation is mainly induced by the scalar isoscalar meson 
$\sigma $ and the vector isoscalar meson $\omega$, while the axial vector current, 
which is more important for the neutrino opacity, is dominantly
affected by the vector isovector meson $\rho $. 

As shown shortly, the structure functions of nucleons are in general 
narrow peaked as a function of the transferred energy in the sense
that their width is much smaller than the transferred 
three momentum. This feature reflects the fact that the nucleon mass is typically  
much greater than the neutrino energy and the scattering is almost isoenergetic. 
If we assume that the scatterings are exactly isoenergetic, that is, 
the structure functions are proportional to the delta function of the 
transferred energy, then we obtain the well known Bruenn's formula \cite{br85}
which we use as a reference in this paper. In some cases this
approximation alone overestimates the cross section
significantly \cite{sc90}. This issue will be addressed again in 
the following sections.

\subsubsection{high density regime}

First we consider the high density regime, $\rho _{b} \geq 2 \times
10^{14}$g/cm$^{3}$. Fig.~\ref{fig1} shows the structure 
functions $R_{1}(k_{0}, |\overrightarrow{\bbox{k}}|)$ and
$R_{2}(k_{0}, |\overrightarrow{\bbox{k}}|)$ 
as a function of the transferred energy $k_{0} $ for the density, 
temperature and proton fraction given in the figure. The transferred 
three momentum $|\overrightarrow{\bbox{k}}|$ is taken to be 
a typical neutrino energy 
$\sim 3 \, T$. The long dashed curves correspond to the structure 
functions of the non-interacting nucleons. The width of the structure
functions gives a measure of average energy exchange between 
neutrino and nucleon. As stated above, we obtain 
the Bruenn's formula \cite{br85} if we neglect this width of these functions
entirely and approximate the structure functions by 
the delta functions of $k_{0}$ as:
\begin{eqnarray}
\label{eq:rapp1}
R_{1}(k_{0} , \, |\overrightarrow{\bbox{k}}|) 
& \approx & 2 \, \pi \, \delta (k_{0} ) 
\, \cdot \, \left[\, {h_{V}^{n}}^{2} \, \eta _{n} \, + \,
{h_{V}^{p}}^{2} \, \eta _{p} \, \right ] \\ 
\label{eq:rapp2}
R_{2}(k_{0} , \, |\overrightarrow{\bbox{k}}|) 
& \approx & 2 \, \pi \, \delta (k_{0} ) 
\, \cdot \, \left[\, {h_{A}^{n}}^{2} \, \eta _{n} \, + \, 
{h_{A}^{p}}^{2} \, \eta _{p} \, \right ] 
\end{eqnarray}
with $\eta _{N} = \displaystyle{\int} \! \frac{2 \, d^{3} \!
 p}{(2\pi)^{3}} \, f_{N}(p) [1 - f_{N}(p)]$ 
for $N = n, p$. Here $f_{N}(p)$ is Fermi-Dirac
 distribution function for the nucleon. 
Note that the long wave length limit $|\overrightarrow{\bbox{k}}| \rightarrow 0$ 
is also taken in this approximation.
It is obvious from the figure that this approximation is not very good 
for this large momentum transfer which is common in the hot proto
neutron star \cite{sc90}. It should be noted that the neglection of the energy 
transfer leads to the overestimation of the reaction rate. 

The short dashed curves are obtained from the first term including
$L_{0}$ on the right hand side of  
Eq.~(\ref{eq:rpaeq}), where only the modified dispersion relation in medium,
that is, the effective mass and potential of nucleons is taken into 
account. Note the effective mass of the nucleon becomes  
as small as $\sim 0.6M_{N}$ in RMF (see \cite{st94a,st94b}). 
The low effective mass renders the structure functions a little bit
wider than for the non-interacting case, since the energy exchange 
between neutrino and nucleon is more facilitated, and lowers 
their amplitudes as well. 
It is important to note that with this modification alone, the neutrino 
scattering rates are considerably reduced \cite{re98a}. However, as shown in the 
preceding section, the inclusion of the effective mass alone 
is not consistent with EOS obtained in RMF.

The solid curves in the figure shows the structure functions with the 
RPA correlation. As already mentioned, the scalar
isoscalar meson $\sigma $ and the vector isoscalar meson 
$\omega $ are the dominant agents of the correlation for $R_{1}(k)$. 
In the high density regime, the contribution of $\omega $ dominates
over that of $\sigma $, making the nuclear force effectively repulsive 
and having the structure function even smaller for small $k_{0}$.
On the other hand, the $\rho $ meson is the dominant
mediator of the nuclear correlation for $R_{2}(k)$. Since the 
typical transferred momentum is still small enough for the short
ranged repulsive force described by the Landau~--~Migdal parameters to 
be dominant, the response of this channel is also suppressed. 
The feature mentioned here is common to the structure functions 
in this density regime. 

As the temperature decreases, the structure function is more skewed 
toward the positive energy transfer $k_{0} > 0$, reflecting the fact that the 
extraction of energy from the nuclear medium becomes more difficult 
due to the Fermi blocking for the down scattered nucleon. This is
shown in Fig.~\ref{fig2}.

On the other hand, as the density becomes higher, the nucleon
effective mass gets smaller ($M^{*}_{N} \sim 0.4 M_{N}$ at $\rho _{b} = 
5\times 10^{14}$g/cm$^{3}$), making the amplitudes of 
the structure functions, both for $R_{1}(k)$ and $R_{2}(k)$, smaller without RPA. 
The $\omega $ meson becomes even more dominant over the 
$\sigma $ meson and thus RPA further reduces the amplitude of $R_{1}(k)$, as
shown in Fig.~\ref{fig3}. 

The case of the proton fraction $Y_{p} = 0.1$ is shown in
Fig.~\ref{fig4}. The effect of RPA increases with decreasing $Y_{p}$,
which could be understood as follows. The vector current correlation comes mainly
from the neutron sector, since the coupling constant for proton 
$h_{V}^{p} \sim 0$. Since the neutron density increases as the proton
fraction decreases, the vector current correlation becomes greater,
reducing the structure function $R_{1}(k)$. 

The total scattering rate of the neutrino with the incident energy 
$E_{\nu }^{in}$ is given by the integration of the structure function 
with respect to the transferred momentum $\overrightarrow{\bbox{k}}$ 
and energy $k_{0} $: 
\begin{eqnarray}
\label{eq:toteq}
R^{tot}(E_{\nu }^{in}) & = & \int \! \frac{d^{3} q_{\nu }^{out}}
{(2 \pi )^{3}} \, \frac{1}{2 E_{\nu }^{in} \, 2 E_{\nu }^{out}} \  
R(E_{\nu }^{in}, \, E_{\nu }^{out}, \, \cos \theta ) 
\  \left [ \, 1 \, - \, f_{\nu }(E_{\nu }^{out}) \, \right ]\nonumber \\
& = & \frac{1}{(2 \pi )^{3}} \, \int ^{\infty}_{0} \! 
2 \pi k \, dk \, \int _{-k}^{k_{0}^{max}} \! \! \! d k_{0} \ 
\frac{E_{\nu }^{in} - k_{0} }{E_{\nu }^{in}} 
\frac{1}{2 E_{\nu }^{in} \, 2 E_{\nu }^{out}} \ 
R(E_{\nu }^{in}, \, E_{\nu }^{out}, \, \cos \theta ) 
\  \left [ \, 1 \, - \, f_{\nu }(E_{\nu }^{out})
\, \right ] ,
\end{eqnarray}
with $k_{0} ^{max} = min(k, \, 2E_{\nu }^{in} - k)$.
If we ignore the energy exchange and the Fermi blocking for the scattered neutrino, 
and insert Eqs.~(\ref{eq:rapp1}) and (\ref{eq:rapp2}) into $R_{1}$ and $R_{2}$ 
in the above equation, we obtain the standard approximate formula~\cite{br85}
\begin{equation}
\label{eq:totap}
R^{tot}(E_{\nu }^{in}) \  \approx \ \sum _{N = n, p} \frac{G_{F}^{2}}{\pi } \ 
\left ( E_{\nu }^{in} \right )^{2} \ 
\left \{ \, \eta _{N} \, \left [ \, \left (h_{V}^{N} \right )^{2} \, + \, 
3 \, \left ( h_{A}^{N} \right )^{2} \, \right ]
\, \right \}
\end{equation}
which is frequently used in literatures. In the following we use this 
rate as a reference in evaluating quantitatively the suppression factors of 
the total scattering rates due to the correlation. It should be noted, however,  
that the difference between these rates also comes from the inclusion of the 
effective nucleon mass and the finite energy and momentum transfer, 
which are neglected in Eq.~(\ref{eq:totap}).

Fig.~\ref{fig5} shows the ratio of $R^{tot}$
evaluated for the first term in Eq.~(\ref{eq:rpaeq}) 
by using the exact formula given in Eq.~(\ref{eq:toteq}) to $R^{tot}$ by
using the approximate expression given in Eq.~(\ref{eq:totap}) as 
a contour map in the density and temperature
plane for the given proton fraction. The contributions from
$R_{1}$ and $R_{2}$ are shown separately. The incident neutrino
energy is assumed to be $3 \, T$, and the blocking factor was dropped
again. 

It is again clear that the reaction rates are substantially smaller than the
standard ones before RPA is included. This is mainly due 
to the decrease of the nucleon effective mass. In fact, as the density 
increases, the reduction becomes greater, which is common to 
the vector current part from $R_{1}$ and the axial vector current part 
from $R_{2}$. The non-monotonic temperature dependence is 
understood as follows: In the standard approximation formula, as
mentioned above, the energy exchange between neutrino and nucleon is
neglected. In the low temperature regime, this neglection tends to
overestimate the Fermi blocking for the scattered nucleon, reducing
the standard rate. As a result, the suppression factor becomes larger
as the temperature increases. For even higher temperature, however,
the width of the structure function cannot be ignored and lowers the
total scattering rate just like the neutrino -- electron scattering. 
Since the latter effect surpasses the former at some temperature, the
suppression factor begins to decrease again above this
temperature, which becomes larger as the density increases.
These features were discussed in the paper by Schinder \cite{sc90} where he 
studied the effect of the recoil of nucleons.

Fig.~\ref{fig6} shows the suppression factor for the reaction rates
with the inclusion of RPA. It is evident that both the vector current and
axial vector current contributions are further suppressed, which is expected
from the above results for the structure functions. Indeed the
reaction rates become less than half the standard rates around the
saturation density. Since the delayed explosion is very sensitive to
the neutrino luminosity and the energy, this suppression could have an 
significant influence on the final outcome of the core collapse. 

we can understand the different density and temperature dependences of 
the vector and axial vector contributions. In the former case, the
RPA suppression is strongly density dependent, since the nuclear force
is determined by the competition of the attractive $\sigma $ meson and 
the repulsive $\omega $ meson. As a result, the reaction rates are
more strongly suppressed in the higher density regime. On the other
hand, the balance between the attractive $\rho $ meson and the
repulsive short range force described by the Landau~--~Migdal
parameter is determined by the momentum transfer,
thus more temperature dependent. Since the nuclear force is more
repulsive for the smaller momentum transfer, that is, in the low
temperature regime, the rate is reduced further there.

Figs.~\ref{fig7} and \ref{fig8} shows the suppression factors for 
the different proton 
fractions. As expected, the axial vector part is more affected, since
the isovector $\rho $ meson is convening its correlation. It is also
clear that the suppression is stronger for the smaller proton fraction.

\subsubsection{low density regime}

The low density regime considered here is characterized by the
appearance of nuclei for low temperatures \cite{sh98,sh99,ms95}. 
Although we cannot treat
the coexistence phase of nuclei and nucleons by the current
formulation and consider here only the gas phase where no nuclei exist, 
the precursor of the phase transition is already imprinted in this regime. 

On the contrary to the previous section, the $\sigma $ meson dominates the 
$\omega $ meson and the isoscalar nuclear force becomes
attractive. As a result, the correlation enhances the reaction rate 
coming from the vector current part. This can be seen in Fig.~\ref{fig9},
where the structure functions $R_{1}$ and $R_{2}$ are shown for the
temperature near the phase boundary. The enhancement of $R_{1}$ 
in the vicinity of $k_{0} = 0$ suggests the existence of the static 
unstable mode \cite{lh89}. In fact, $R_{1}$ diverges on the so-called spinodal
line \cite{hp93}. From the thermodynamic consistency discussed above, 
$\left ( \frac{\partial N}{\partial \mu } \right )_{T}$ also diverges
on this line. This behavior is better indicated in Fig.~\ref{fig10},
in which the enhancement factor with respect to the standard Bruenn's
formula is plotted as a contour in the density temperature plane. As
the temperature decreases, the reaction rate becomes larger. The lower 
boundary of the computation region roughly corresponds to the spinodal line
obtained by the current model. Incidentally, the axial vector
contribution is hardly affected since the $\sigma $ and $\omega $
mesons are minor contributors for this channel. 

Since in reality this liquid-gas phase transition occurs not by the
spinodal decomposition but by the nucleation, the actual phase 
boundary does not correspond to the spinodal line except for the
critical point, and the spinodal line is never reached 
\cite{sh98,sh99,ms95}. The critical
temperature obtained from RPA is about $15$MeV, which is actually
consistent with the value obtained from EOS by Shen et al. \cite{sh98,sh99}.

Although the enhancement factor is still relatively large in the
region not very close to the phase boundary, this effect has only  
minor importance for the neutrino opacity, because the vector
current contribution is typically much smaller ($\sim 1/4$) than the axial
vector contribution. Provided that the axial vector part is slightly suppressed 
in this regime due to the repulsive nuclear interaction for that
channel, the total scattering rate will not be increased significantly 
unless the critical point is approached very closely.

\subsection{Charged Current Reactions}

The RPA correlation is calculated for the charged current
reactions just the same way as for the neutral current reactions 
except that the vector isovector $\rho $ meson is the only mediator of the
correlation. It is noted again that since only the spin-transverse correlation 
matters for the neutrino reaction rates in RPA, the pion does not
make any contribution. As a result, for relatively small momentum
transfer of our current interest the nuclear force is repulsive and 
the correlation always tends to suppress both the vector current and 
the axial vector current contribution to 
the charged current reaction rates like the axial vector part of
the neutral current. This is true not only for the high density
regime but also for the low density regime
unlike the vector current part of the neutral current reaction rates
which is enhanced by the liquid -- gas phase transition.

Fig.~\ref{fig11} shows a typical modification of the structure functions 
around the saturation density for the symmetric nuclear
matter. Depicted are $R_{1}(k_{0}, |\overrightarrow{\bbox{k}}|)$ 
and $R_{2}(k_{0}, |\overrightarrow{\bbox{k}}|)$ for the $e^{-} + p \rightarrow 
\nu _{e} + n$ reaction
with the outgoing neutrino energy of $\sim \mu _{e}$, the electron
chemical potential. It is clear again that the decrease of the nucleon 
effective mass alone (the short dashed curve) accounts for 
more than half of the suppression with respect to the non-interacting
case (the long dashed curve).  Moreover, as 
the electron chemical potential is quite large, $\sim 250 $MeV, 
the average energy transfer from the incident electron to the nucleon
becomes also large. Hence the complete neglection of energy transfer
assumed in the approximate formula \cite{br85} leads to sizable overestimation of 
the reaction rates for this high energy neutrino emission. 

The results change qualitatively for the asymmetric matter, since
the potential for the neutron is different from that for the proton 
in medium. This is understood as follows. In the non-relativistic 
limit the dispersion relation for the nucleon is written as
\begin{equation}
E_{N} (\overrightarrow{\bbox{k}}) \ = \ \frac{|\overrightarrow{\bbox{k}}|^{2}}
{2 M_{N}^{*}} \, + \, U^{pot}_{N} \quad ,
\end{equation}
and the potential difference is given in RMF by
\begin{equation}
\Delta U_{pn}^{pot} \ \equiv \  U^{pot}_{p} \, - \, U^{pot}_{n} \ = \ 
2 \, g_{\rho } \langle \rho \rangle \ = \ 2\, \frac{g_{\rho }^{2}}
{m_{\rho }^{2}} \, (n_{p} - n_{n}) \quad .
\end{equation}
Thus, in neutron rich matter, the proton is more strongly bound than the 
neutron. This difference of potentials serves as a threshold for
the $p \rightarrow n$ reaction as the mass difference of nucleons does for the
reaction in vacuum. It is, however, ignored in the 
approximate formula \cite{br85}, which is given by  
\begin{eqnarray}
\label{eq:brapc1}
R_{1}(k_{0} , \, |\overrightarrow{\bbox{k}}|) 
& \approx & 2 \, \pi \, \delta (k_{0} - \Delta) 
\, \cdot \, {g_{V}}^{2} \  \eta _{pn} \\
\label{eq:brapc2}
R_{2}(k_{0} , \, |\overrightarrow{\bbox{k}}|) 
& \approx & 2 \, \pi \, \delta (k_{0} - \Delta) 
\, \cdot \, {g_{A}}^{2} \  \eta _{pn} 
\end{eqnarray}
with $\eta _{pn} = \displaystyle{\int} \! \frac{2 \, d^{3} \! p}{(2\pi)^{3}} 
\, f_{p}(p) [1 - f_{n}(p)]$.  
Hence the structure functions in this approximation are  delta functions
placed at $k_{0} = \Delta \equiv M_{n} - M_{p}$.  

With the potential difference taken into account, however, the
structure functions are shifted by $ - \Delta U^{pot}_{pn}$, thus they
do not agree with the approximate ones even in the static and long wave length
limit with $M_{N}^{*} \rightarrow M_{N}$. 
This is seen in Fig.~\ref{fig12}, where the structure functions for
the asymmetric matter ($Y_{p} = 0.3$) 
are shown for the non-interacting case (the long dashed curve) 
as well as the case with the
nucleon effective mass and potential alone included (the short dashed
curve) and the case with RPA also included (the solid curve). 
In the non-interacting case, the structure functions are 
located around $k_{0} \sim \Delta$, while in the other cases they are
moved toward the positive energy transfer as stated above. 

The total $\nu _{e}$ emission rate is given for an outgoing neutrino
energy $E_{\nu }^{out}$ as
\begin{eqnarray}
\label{eq:toteqc}
R^{tot}(E_{\nu }^{out}) & = & \int \! \frac{d^{3} q_{e}^{in}}
{(2 \pi )^{3}} \, \frac{1}{2 E_{e}^{in} \, 2 E_{\nu }^{out}} \  
R(E_{e}^{in}, \, E_{\nu }^{out}, \, \cos \theta ) 
\  f_{e}(E_{e}^{in}) \nonumber \\
& = & \frac{1}{(2 \pi )^{3}} \, \int ^{\infty}_{0} \! 
2 \pi k \, dk \, \int _{k_{0}^{-}}^{k_{0}^{+}} \! \! \! d k_{0} \ 
\frac{E_{\nu }^{out} + k_{0} }{E_{\nu }^{out}} 
\frac{1}{2 E_{e}^{in} \, 2 E_{\nu }^{out}} \ 
R(E_{e}^{in}, \, E_{\nu }^{out}, \, \cos \theta ) 
\  f_{e}(E_{e}^{in})
\quad,
\end{eqnarray}
where $k_{0} ^{\pm} = \sqrt{m_{e}^{2} + (E_{\nu }^{out} \pm k)^{2}} - 
E_{\nu }^{out}$. Inserting Eqs.~(\ref{eq:brapc1}), (\ref{eq:brapc2}) 
into the above equation,  we obtain the standard approximate formula:
\begin{equation}
\label{eq:totapc}
R^{tot}(E_{\nu }^{out}) \ \approx \  
\frac{G_{F}^{2}}{\pi } \  ({g_{V}}^{2} + \, 3 \, {g_{A}}^{2}) \  \eta _{pn}
\ (E_{\nu }^{out} + \Delta )^{2} \, \sqrt{1 - 
\left (\frac{m_{e}}{E_{\nu }^{out} + \Delta } \right )^{2}} \ 
f_{e}(E_{\nu }^{out} + \Delta) \quad .
\end{equation}
For the symmetric nuclear matter, the density and temperature
dependence of the suppression factor of the total emission rates, that 
is the ratio of Eq.~(\ref{eq:toteqc}) to Eq.~(\ref{eq:totapc}), 
is essentially the same as that of the axial vector contribution to  
the neutral current reaction. For the asymmetric matter, as understood from 
Eq.~(\ref{eq:toteqc}), the $k$ integration has greater contribution
from larger momentum transfer due to the shift of the 
structure functions, which tends to get the reaction rates larger
thanks to the larger phase volume. However, as $E_{\nu }^{out}$
comes closer to $\mu _{e}$, the electron population
starts to deplete $f_{e}(E_{e}^{in}) \lesssim 1$ and, as a result the emission rate
becomes smaller. This occurs for smaller $E_{\nu }^{out}$ 
in Eq.~(\ref{eq:toteqc}) than in
Eq.~(\ref{eq:totapc}) due to the large width of the structure
function. It follows that the total emission rate is greater
than the approximate one given by Eq.~(\ref{eq:totapc}) for 
neutrinos with energy $E_{\nu }^{out} \ll \mu _{e}$ and smaller 
for neutrinos with $E_{\nu }^{out} \lesssim \mu _{e}$. This is indicated in 
Fig.~\ref{fig13}. Thus lower energy electron neutrinos are
produced more efficiently than predicted from the standard approximate
formula. It is also seen that the RPA correction 
is not very large since the nuclear force becomes less repulsive for
the large momentum transfer considered here.

Fig.~\ref{fig14} displays the structure functions for a lower
temperature. It is evident that their widths became smaller. This is
because nucleons become more degenerate as the temperature is lowered
and the large energy transfer is required to overcome the difference
of chemical potentials between neutron and proton.  
The comparison of Fig.~\ref{fig15}, which represents the emission
rates for the same temperature, with Fig.~\ref{fig13} shows that 
the emission of high energy neutrinos is severely suppressed for the
low temperature, because it is contributed from the high
energy tail of degenerate electrons. It is also seen
from the figure that the approximate formula tends 
to overestimate the Fermi blocking factor for the outgoing neutron by 
neglecting the finite energy transfer, thus underestimating the 
emission rate. The RPA correction is even smaller in this case, since 
the narrow  structure functions favor higher momentum transfer. 

As the density is increased or the proton fraction is decreased, the
potential difference becomes larger. Hence the features mentioned 
above become more prominent, which is demonstrated in Figs.~\ref{fig16}
and \ref{fig17}. In the latter case, since the electron chemical
potential is also decreased, it is possible for small $Y_{p}$ and $T$
that the structure functions and the electron distribution function 
in the integrand of Eq.~(\ref{eq:toteqc}) do not give a significant
overlap and the emissivity is strongly reduced. This is well known 
suppression of the URCA process \cite{st83}. 

The structure functions for the $n \rightarrow p$ reactions are derived from
those for the $p \rightarrow n$ reactions by using the 
detailed balance relation expressed as 
\begin{equation}
\label{eq:baleq}
R^{(n, \, p)}(k) \ = \ e ^{\beta (k_{0} - \Delta \mu _{pn})} 
\ R^{(p, \, n)}(-k) \quad ,
\end{equation}
where $\Delta \mu _{pn} = \mu _{p} - \mu _{n}$ is the difference of
the nucleon chemical potentials. In this paper, however, the $n
\rightarrow p$
structure functions were calculated in the same way as the $p
\rightarrow n$
counterparts, and we confirmed directly that the above relation is actually 
satisfied. 

As expected, the structure functions for the $n \rightarrow p$ reactions 
are shifted toward the negative energy
transfer due to the potential difference for  proton and neutron
in asymmetric matter, which is shown in Fig.~\ref{fig18}. The RPA
correction accounts for about half of the total suppression in this
case. The total emissivity
of $\overline{\nu }_{e} $ via $e^{+} + n \rightarrow \overline{\nu
}_{e} + p$ is
obtained from Eq.~(\ref{eq:toteqc}) with the following exchanges: 
$p \leftrightarrow n$, $e \leftrightarrow e^{+}$ and $\nu _{e} 
\leftrightarrow \overline{\nu }_{e}$. The important difference 
from the $\nu _{e}$ case comes from the fact that positrons are not 
degenerate in the supernova core. The shift of the structure 
functions in the direction of the negative energy transfer gives greater 
weights to positrons with lower energies which are more populated, 
leading to larger emissivity of $\overline{\nu }_{e}$. On the other 
hand, the RPA correction tends to reduce the rate as in the $\nu _{e}$ 
case. These trends are shown in Fig.~\ref{fig19}, where the
$\overline{\nu }_{e}$ emissivity is plotted as a function of the
energy of the emitted neutrino. The potential difference for 
neutron and proton provides the emitted neutrino with certain 
energy ($\sim 30$MeV in this case). 
Although the correlation reduces the emissivity substantially, 
the emissivity of high energy neutrinos are larger even after RPA 
is included than for the approximate formula.

Thus far, we discussed the $\nu $ emissivities. The corresponding 
absorption rates of $\nu _{e}$ via $\nu _{e} + n \rightarrow e + p$ and 
$\overline{\nu }_{e} $ via $\overline{\nu }_{e} + p \rightarrow e^{+}
+ n$ are obtained by the detailed balance equation:
\begin{equation}
R^{tot}_{ab}(E_{\nu }) \ = \ e ^{\beta (E_{\nu } - \mu _{\nu } )} \ 
R^{tot}_{em}(E_{\nu }),
\end{equation}
where the neutrino chemical potential is defined as 
$\mu _{\nu _{e}} \equiv \mu _{e} + \mu _{p} - \mu _{n}$ for $\nu _{e}$ 
and $\mu _{\overline{\nu }_{e}} = - \mu _{\nu _{e}}$ for $\overline {\nu }_{e}$. 
This relation is easily proven from Eq.~(\ref{eq:baleq}). 

\section{Conclusion}

In view of the application to supernova simulations 
\cite{sy97,sy99}, we have calculated the
neutrino -- nucleon reaction rates in hot and dense supernova cores with the RPA
correlations taken into account. The approximations are based on the
Lagrangian used in  RMF for the nuclear EOS, 
where the nonlinear $\sigma $ and $\omega $ meson terms are
included. This ensures the thermodynamic consistency in the static and long wave
length limit between the EOS we use in the supernova simulations and
the vector current correlation of nucleons which gives partial
contributions to the neutrino -- nucleon scattering rates. The same
method is extended to the axial vector current as well as to the isovector
current correlations with the residual interactions added which are
described by the tensor coupling of the $\rho $ meson and the
so-called Landau-Migdal parameters. 

We have found that the neutral current reactions are suppressed
substantially in the high density regime, $\rho \gtrsim
10^{14}$g/cm$^{3}$, due to the repulsive nature of the nuclear
forces. In the lower density regime, however, the vector current contributions
are enhanced as the temperature is decreased and the boundary of the
gas phase, which is placed by RPA consistently with EOS, 
is approached. Although this is interesting itself, the total
scattering rates are not enhanced significantly except for the very
vicinity of the critical point, since the axial vector
contribution is dominant over the vector current one and 
is unaffected by the liquid -- gas phase transition. 
We have also shown that the
neglection of the finite energy exchange between neutrino and 
nucleon, which is assumed in the standard approximation formula, 
can lead to a significant overestimation of the total reaction rates in
the high temperature regime where the typical incident neutrino energy 
is no longer much smaller than the nucleon effective mass. 

In the case of the charged current reactions, it is more important to
take proper account of the potential difference for neutron and
proton than to consider the RPA correlation, since it serves as a
threshold of the reaction just as a mass difference does in vacuum. In 
fact, with the potential difference included the $\nu _{e}$ emissivity 
via $e + p \rightarrow \nu _{e} + n$ is much smaller for
higher energy neutrinos than expected from the standard formula. 
On the other hand, the emission of low energy antineutrinos via 
$e^{+} + n \rightarrow \overline{\nu }_{e} + p$ is reduced. 
In both cases, the RPA correlation
reduces the total rates further. 

These modifications of the neutrino -- nucleon reaction rates could
have an important impact on the mechanism of collapse-driven
supernovae, since they are the dominant opacity sources for
neutrinos. The smaller opacity leads to the larger neutrino
luminosity, hence to the greater heating rate behind the stagnated
shock wave, which gives a better odds to get a successful
explosion. The changes of neutrino mean free paths, on the other hand, 
could also affect the strength and region of convections in the proto
neutron star, which is one of the key ingredients to determine the
neutrino luminosity and energy. These issues will be discussed in
detail elsewhere \cite{ja99}. 

It is clear that RPA does not include everything. As understood from
the fact that RPA is consistent with the mean field theory for the
equation of state, RPA studies the response of the mean field to the
external disturbance. The additional correlations might be 
induced by some reactions such as collisions of two nucleons via spin and isospin
dependent reactions, as pointed out by some authors. Although these effects can in
principle be included in the above formulation by taking the corresponding
higher order corrections to the self-energy, it is extremely difficult 
to achieve it in practice. In this paper we employed RPA respecting the
consistency between the calculated reaction rates and EOS we have
currently at our disposal. Nevertheless, these issues are remaining to be studied
further. 

\acknowledgments

One of the authors (S.Y.) greatfully acknowledges critical discussions
with H.-Th. Janka and G. Raffelt. He would highly appreciate various
support by E. M\"{u}ller during his stay in Max Planck Institut
f\"{u}r Astrophysik. This work is partially supported by the 
Grants-in-Aid for the Center-of-Excellence (COE) Research of
the Ministry of Education, Science, Sports and Culture of Japan to
RESCEU (No.07CE2002).

\appendix
\section*{causal Green functions}

Once the formulation is established in the imaginary time formalism, it is more 
convenient to work with quantities which are analytically continued from the 
imaginary time domian to the real time domain. Then
Eq.~(\ref{eq:rpaeq}) is still valid with the response functions 
replaced with the retarded Green functions. These real 
time retarded Green functions Eq.~(\ref{eq:defret}) are derived 
from the corresponding causal Green functions defined by 
Eq.~(\ref{eq:defcau}) via the relation Eq.~(\ref{eq:disp}).
Thus all we have to do is to calculate $\Pi ^{R0}_{N}$ and then to solve 
Eq.~(\ref{eq:rpaeq}) which is now rewritten in momentum space as
\begin{equation}
{\Pi _{N}^{R}}^{a b}(k) \  =  \ {\Pi _{N}^{R0}}^{a b}(k) \ 
- \  \sum _{c} \,
{\Pi ^{R0}_{N}}^{a c}(k) \, \cdot \, V^{c}_{pot}(k) \, \cdot \, 
{\Pi _{N}^{R}}^{c b}(k) \quad .
\end{equation}
Here ${\Pi _{N}^{R0}}^{a b}(k) $ is derived by Eq.~(\ref{eq:disp}) from 
${\Pi _{N}^{0}}^{a b}(k) $ which is expressed as
\begin{equation}
\label{eq:intpi}
i \, {\Pi _{N}^{0}}^{a b}(k) \ = \ Tr \int \! \frac{d^{4}q}{(2 \pi)^{4}}
\, \Gamma ^{a} \, G(q + k) \, \Gamma ^{b} \, G(q) \quad ,
\end{equation}
where $\Gamma $'s stand for $\bbox{1}, \gamma ^{\mu}, \sigma ^{\mu \nu }, 
\gamma ^{\mu} \gamma _{5}$. $G(k)$ is a single particle Green function of nucleon 
given by RMF: 
\begin{eqnarray}
G(k) & = & G_{F}(k) \ + \ G_{D}(k) \\
G_{F}(k) & = & \frac{\ovslash{\bar{k}} + M_{N}^{*}}
{\bar{k}^{2} - {M_{N}^{*}}^{2} + i \varepsilon} \\
G_{D}(k) & = & (\ovslash{\bar{k}} + M_{N}^{*})  \cdot 
2 \pi i \, \delta (\bar{k}^{2} - {M_{N}^{*}}^{2})
\left \{ \Theta (\bar{k}_{0}) \, f_{N}(k_{0}) \, + \, 
\Theta (- \bar{k}_{0}) \, f_{\bar{N}}(- k_{0})
\right \} \quad ,
\end{eqnarray}
where $\bar{k} = (k_{0} - \Sigma _{V}, \bbox{k})$, $M_{N}^{*} =
M_{N} + \Sigma _{S}$ with $\Sigma _{S}, \Sigma _{V}$ the sclar and 
vector nucleon self-energies given by RMF, and $f_{N}, f_{\bar{N}}$ 
are Fermi distribution functions for nucleon and antinucleon,
respectively. Note that since the above 
Green function $G(k)$ already includes the effective mass and potential, 
$\Pi _{N}^{0}(k)$ is different from the response function 
of non-interacting nucleons. As shown in the above equation, the Green 
function is decomposed into two contributions, that is, 
the Feynman part $G_{F}(k)$ and the density-dependent part
$G_{D}(k)$. As a usual practice, we ignore the vacuum polarization 
coming from the Feynman part alone in calculating $\Pi _{N}^{0}(k)$.  
Taking the proper linear combination of ${\Pi _{N}^{R}}^{ab}$,
we obtain the retarded Green function for the nucleon weak current, from which 
using again Eq.~(\ref{eq:disp}), we finally arrive at the structure 
functions of nucleons. The explicit expressions of
Eq.~(\ref{eq:intpi}) are found, for example, in~\cite{re98c,hw91,sm89}.

\begin{figure}
\caption{The vector and axial vector structure functions 
$R_{1}(k_{0}, \, |\protect\overrightarrow{\protect\bbox{k}}|)$ and 
$R_{2}(k_{0}, \, |\protect\overrightarrow{\protect\bbox{k}}|)$ for the neutral
current as functions of the transfer energy $k_{0}$. 
The density $\rho _{b}$, temperature $T$, proton fraction $Y_{p}$ 
and the absolute value of the transferred three momentum 
$|\protect\overrightarrow{\protect\bbox{k}}|$  
are shown in the figure. The long dashed curve represents the
non-interacting case, while the short dashed curve shows the case for 
the effective mass of the nucleon being taken
into account. The solid curve shows the
results of RPA.}
\label{fig1}
\end{figure}

\begin{figure}
\caption{The vector and axial vector structure functions 
$R_{1}(k_{0} , \, |\protect\overrightarrow{\protect\bbox{k}}|)$ and 
$R_{2}(k_{0} , \, |\protect\overrightarrow{\protect\bbox{k}}|)$ for the neutral
current as functions of the transfer energy $k_{0}$. 
The density $\rho _{b}$ and proton fraction $Y_{p}$ are the same
as in Fig.~\protect\ref{fig1}, but the temperature $T$ is smaller in
this case, and the transferred three momentum 
$|\protect\overrightarrow{\protect\bbox{k}}|$ is scaled as 
$\sim 3 \, T$. The explicit values are shown in the figure. 
}
\label{fig2}
\end{figure}

\begin{figure}
\caption{The vector and axial vector structure functions 
$R_{1}(k_{0} , \, |\protect\overrightarrow{\protect\bbox{k}}|)$ and 
$R_{2}(k_{0} , \, |\protect\overrightarrow{\protect\bbox{k}}|)$ for the neutral
current as functions of the transfer energy $k_{0}$. 
The parameters are the same as 
in Fig.~\protect\ref{fig1} but for the higher density, $\rho _{b} = 5
\times 10^{14}$g/cm$^{3}$ as indicated in the figure.}
\label{fig3}
\end{figure}

\begin{figure}
\caption{The vector and axial vector structure functions 
$R_{1}(k_{0} , \, |\protect\overrightarrow{\protect\bbox{k}}|)$ and 
$R_{2}(k_{0} , \, |\protect\overrightarrow{\protect\bbox{k}}|)$ for the neutral
current as functions of the transfer energy $k_{0}$. 
The parameters are the same as Fig.~\protect\ref{fig1} but for the different proton
fraction $Y_{p} = 0.1$. 
}
\label{fig4}
\end{figure}

\begin{figure}
\caption{The suppression factors for the total scattering rates,
$R^{tot}(E_{\nu }^{in})$ defined in Eq.~(\protect\ref{eq:toteq}), from the
expressions of Bruenn for vector currents (upper figure)
and for axial vector currents (lower figure). Here $E_{\nu }^{in}$ is
taken as $E_{\nu }^{in} = 3 T$. The proton fraction is chosen as
$Y_{p} = 0.3$. The suppressions are caused by the
effective mass for nucleons and the effect of the transfer energy
spectra instead of the $\delta $-function.
}
\label{fig5}
\end{figure}

\begin{figure}
\caption{The suppression factors for the total scattering rates,
$R^{tot}(E_{\nu }^{in})$ defined in Eq.~(\protect\ref{eq:toteq}), from the
expressions of Bruenn for vector currents (upper figure)
and for axial vector currents (lower figure). $E_{\nu }^{in} = 3 T$ and
$Y_{p} = 0.3$ are used here also as in Fig.~\protect\ref{fig5}. In
this case, the 
suppressions are caused by the RPA correlations in 
addition to the effects considered in Fig.~\protect\ref{fig5}.
}
\label{fig6}
\end{figure}

\begin{figure}
\caption{The suppression factors for the total scattering rates,
$R^{tot}(E_{\nu }^{in})$ defined in Eq.~(\protect\ref{eq:toteq}), from the
expressions of Bruenn for vector currents (upper figure)
and for axial vector currents (lower figure). The conditions are the
same as those in Fig.~\protect\ref{fig6}, but the proton fraction is reduced
here to $Y_{p} = 0.1$.
}
\label{fig7}
\end{figure}

\begin{figure}
\caption{The suppression factors for the total scattering rates,
$R^{tot}(E_{\nu }^{in})$ defined in Eq.~(\protect\ref{eq:toteq}), from the
expressions of Bruenn for vector currents (upper figure)
and for axial vector currents (lower figure). The conditions are the
same as those in Fig.~\protect\ref{fig6}, but the proton fraction is changed
here to $Y_{p} = 0.5$.
}
\label{fig8}
\end{figure}

\begin{figure}
\caption{The structure functions 
$R_{1}(k_{0} , \, |\protect\overrightarrow{\protect\bbox{k}}|)$ and 
$R_{2}(k_{0} , \, |\protect\overrightarrow{\protect\bbox{k}}|)$ for the neutral
current in the low density regime. The density $\rho _{b}$, 
temperature $T$, proton fraction $Y_{p}$ 
and the absolute value of the transferred three momentum
$|\protect\overrightarrow{\protect\bbox{k}}|$  
are shown in the figure. Only the effective mass of the nucleon is taken
into account for the short dashed curve, while RPA is included for the
solid curve.}
\label{fig9}
\end{figure}

\begin{figure}
\caption{The enhancement factor for the vector current contribution
(upper) and 
the suppression factor for the axial vector current part (lower) for 
$R^{tot}(E_{\nu }^{in})$ defined in Eq.~(\protect\ref{eq:toteq}) are shown for 
the low density regime. Here the proton fraction is $Y_{p} = 0.3$. 
The dark regions are outside the
computational grid and roughly correspond to the coexistence phase of
nucleons and nuclei, where a special care is required to treat the
non-uniform matter.}
\label{fig10}
\end{figure}

\begin{figure}
\caption{The structure functions 
$R_{1}(k_{0} , \, |\protect\overrightarrow{\protect\bbox{k}}|)$ and 
$R_{2}(k_{0} , \, |\protect\overrightarrow{\protect\bbox{k}}|)$ for the charged 
current reactions, where protons change into neutrons. The density $\rho _{b}$, 
temperature $T$, proton fraction $Y_{p}$ 
and the absolute value of the transferred three momentum
$|\protect\overrightarrow{\protect\bbox{k}}|$  
are shown in the figure. The long dashed curve represents the
non-interacting case while the short dashed curve shows the case for 
the effective mass of the nucleon being taken
into account. The solid curve shows the
results of RPA.}
\label{fig11}
\end{figure}

\begin{figure}
\caption{The structure functions 
$R_{1}(k_{0} , \, |\protect\overrightarrow{\protect\bbox{k}}|)$ and 
$R_{2}(k_{0} , \, |\protect\overrightarrow{\protect\bbox{k}}|)$ for the charged 
current reactions, where protons change into neutrons, for the proton
fraction, $Y_{p} = 0.3$.
The three momentum transfer
$|\protect\overrightarrow{\protect\bbox{k}}|$ 
is also changed to the electron chemical potential.
The notations for curves are the same as in Fig.~\protect\ref{fig11}.}
\label{fig12}
\end{figure}

\begin{figure}
\caption{The total $\nu _{e}$ emission rate defined in 
Eq.~(\protect\ref{eq:toteqc}) as a function of the 
emitted neutrino energy $E_{\nu }^{out}$. The upper panel shows the
rate coming from the vector current and the lower panel represents 
the axial vector current contribution. The long dashed curves are
obtained from the Bruenn's approximation formula, while the short dashed
curves represent the results with the nucleon effective mass and 
the isovector potential being included. The solid curves 
correpond to the RPA results.} 
\label{fig13}
\end{figure}

\begin{figure}
\caption{The structure functions 
$R_{1}(k_{0} , \, |\protect\overrightarrow{\protect\bbox{k}}|)$ and 
$R_{2}(k_{0} , \, |\protect\overrightarrow{\protect\bbox{k}}|)$ for the charged 
current reactions, where protons change into neutrons. The density $\rho _{b}$, 
temperature $T$, proton fraction $Y_{p}$ 
and the absolute value of the transferred three momentum
$|\protect\overrightarrow{\protect\bbox{k}}|$  
are shown in the figure. The long dashed curve represents the
non-interacting case while the short dashed curve shows the case for 
the effective mass of the nucleon being taken
into account. The solid curve shows the
results of RPA. Here the temperature is $T = 10$MeV.
The three momentum transfer
$|\protect\overrightarrow{\protect\bbox{k}}|$ 
is also changed to the electron chemical potential.
The notations for curves are the same as in Fig.~\protect\ref{fig11}.}
\label{fig14}
\end{figure}

\begin{figure}
\caption{The total $\nu _{e}$ emission rate defined in 
Eq.~(\protect\ref{eq:toteqc}) as a function of the 
emitted neutrino energy $E_{\nu }^{out}$. The upper panel shows the
rate coming from the vector current and the lower panel represents 
the axial vector current contribution. The long dashed curves are
obtained from the Bruenn's approximation formula, while the short dashed
curves represent the results with the nucleon effective mass and 
the isovector potential being included. The solid curves 
correpond to the RPA results. Here the temperature is $T = 10$MeV. } 
\label{fig15}
\end{figure}

\begin{figure}
\caption{The total $\nu _{e}$ emission rate defined in 
Eq.~(\protect\ref{eq:toteqc}) as a function of the 
emitted neutrino energy $E_{\nu }^{out}$. The upper panel shows the
rate coming from the vector current and the lower panel represents 
the axial vector current contribution. The long dashed curves are
obtained from the Bruenn's approximation formula, while the short dashed
curves represent the results with the nucleon effective mass and 
the isovector potential being included. The solid curves 
correpond to the RPA results. Here the density is $\rho _{b} = 
5 \times 10^{14}$g/cm$^{3}$} 
\label{fig16}
\end{figure}

\begin{figure}
\caption{The total $\nu _{e}$ emission rate defined in 
Eq.~(\protect\ref{eq:toteqc}) as a function of the 
emitted neutrino energy $E_{\nu }^{out}$. The upper panel shows the
rate coming from the vector current and the lower panel represents 
the axial vector current contribution. The long dashed curves are
obtained from the Bruenn's approximation formula, while the short dashed
curves represent the results with the nucleon effective mass and 
the isovector potential being included. The solid curves 
correpond to the RPA results. Here the proton fraction is $Y_{p} = 0.1$. } 
\label{fig17}
\end{figure}

\begin{figure}
\caption{The structure functions $R_{1}(k_{0} , \, |
\protect\overrightarrow{\protect\bbox{k}}|)$ and 
$R_{2}(k_{0} , \, |\protect\overrightarrow{\protect\bbox{k}}|)$ for the charged 
current reactions, where neutrons change into protons. The density $\rho _{b}$, 
temperature $T$, proton fraction $Y_{p}$ 
and the absolute value of the transferred three momentum $|\protect\bbox{k}|$  
are shown in the figure. The long dashed curve represents the
non-interacting case while the short dashed curve shows the case for 
the effective mass and the isovector potential of the nucleon being taken
into account. The solid curve shows the
results of RPA.}
\label{fig18}
\end{figure}

\begin{figure}
\caption{The total $\overline{\nu }_{e}$ emission rate as a function of the 
emitted neutrino energy $E_{\nu }^{out}$. The upper panel shows the
rate coming from the vector current and the lower panel represents 
the axial vector current contribution. The long dashed curves are
obtained from the Bruenn's approximation formula, the short dashed
curves represent the results with the nucleon effective mass and the isovector 
potential being included. The solid curves display the RPA results.} 
\label{fig19}
\end{figure}

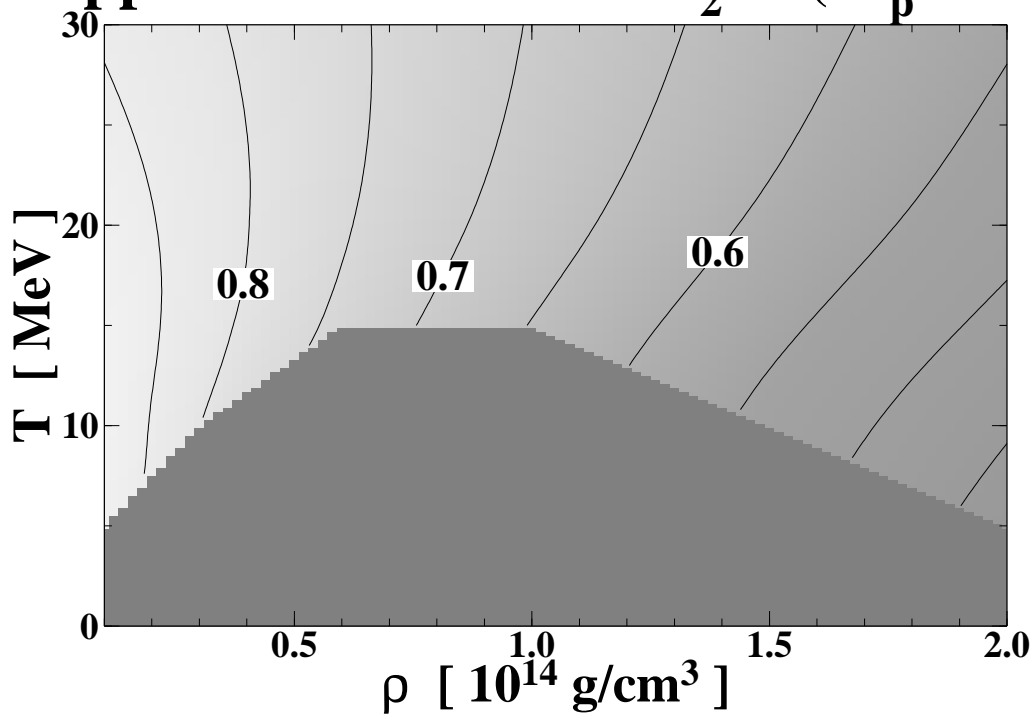
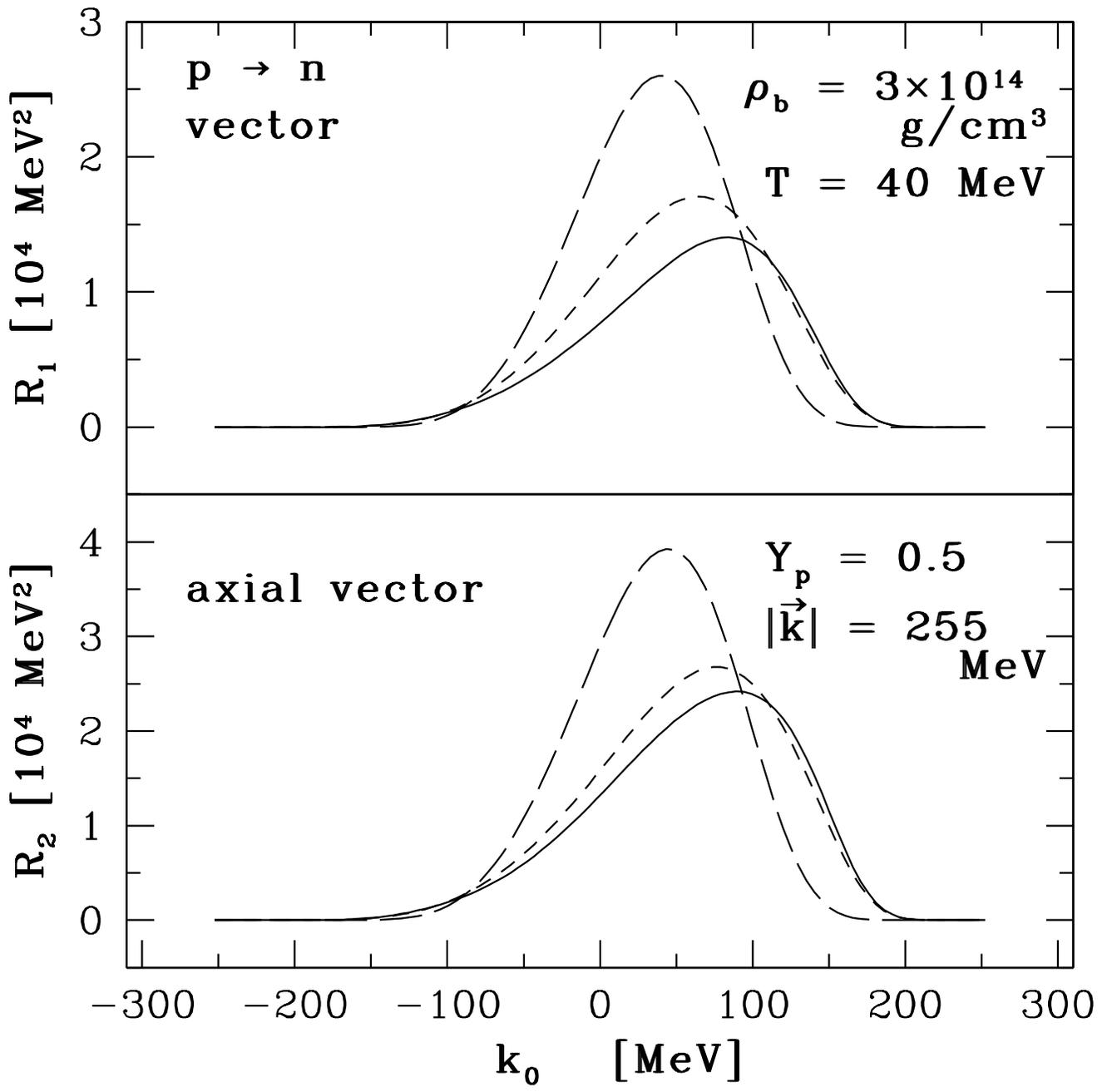
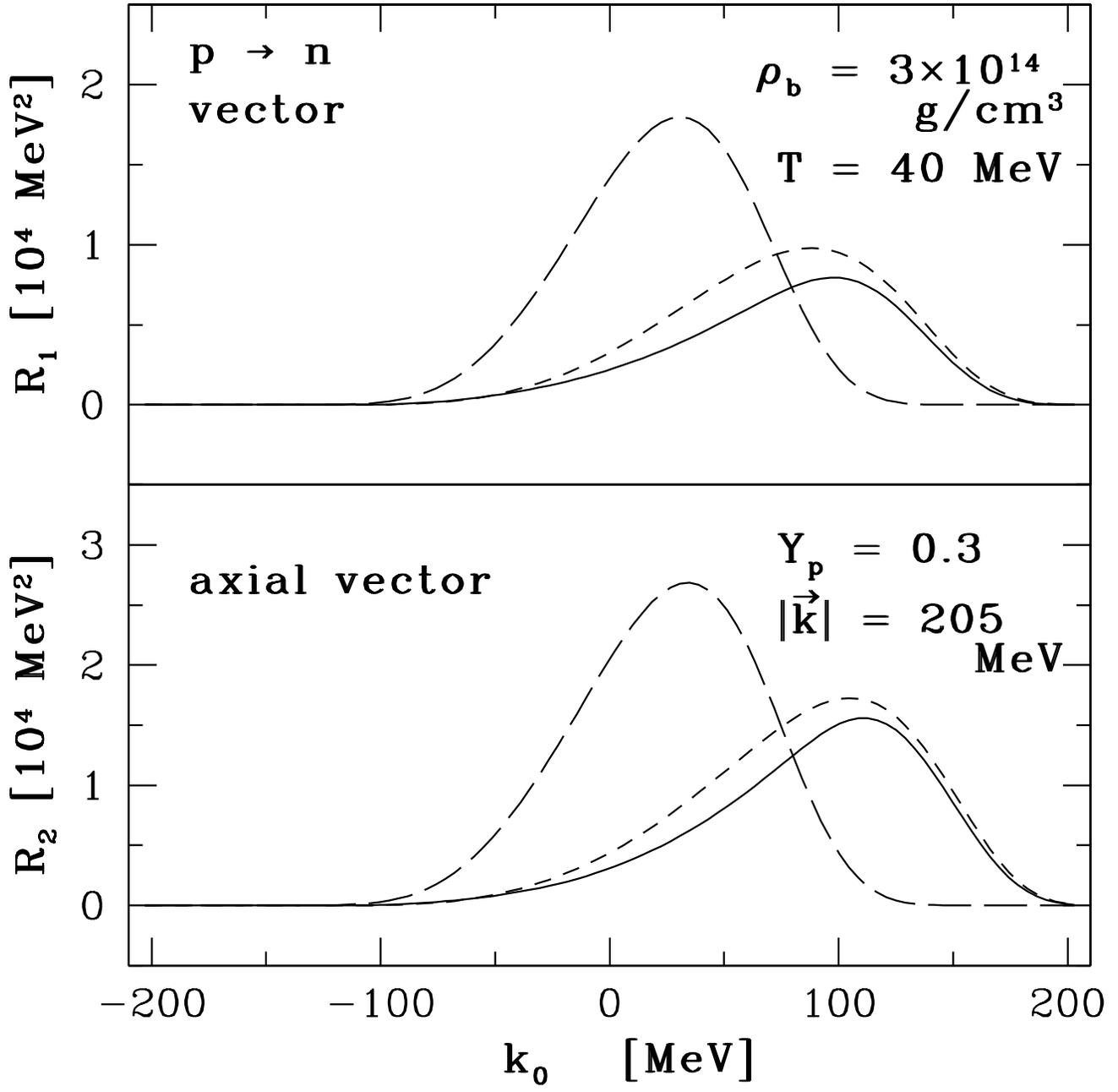
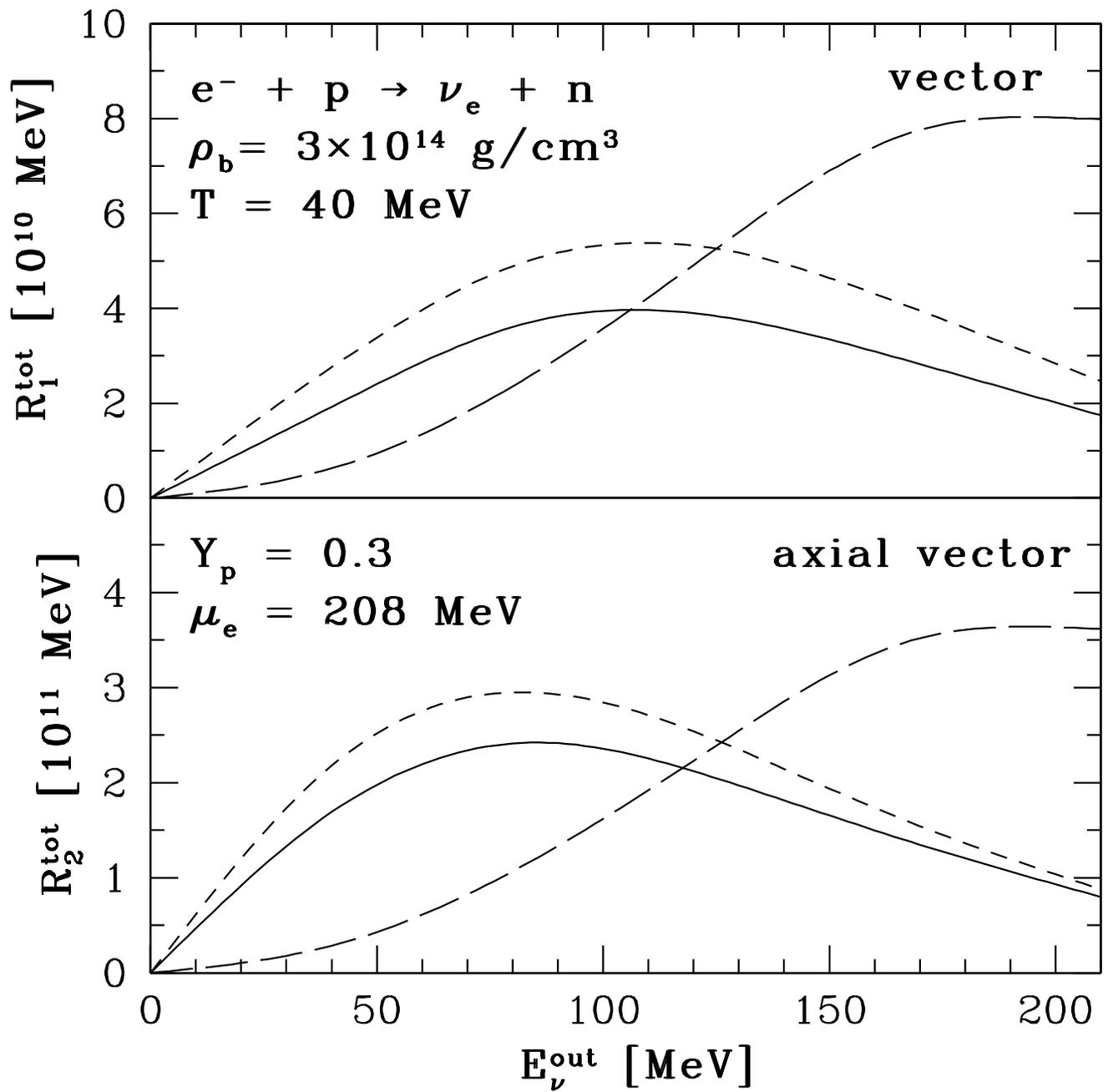
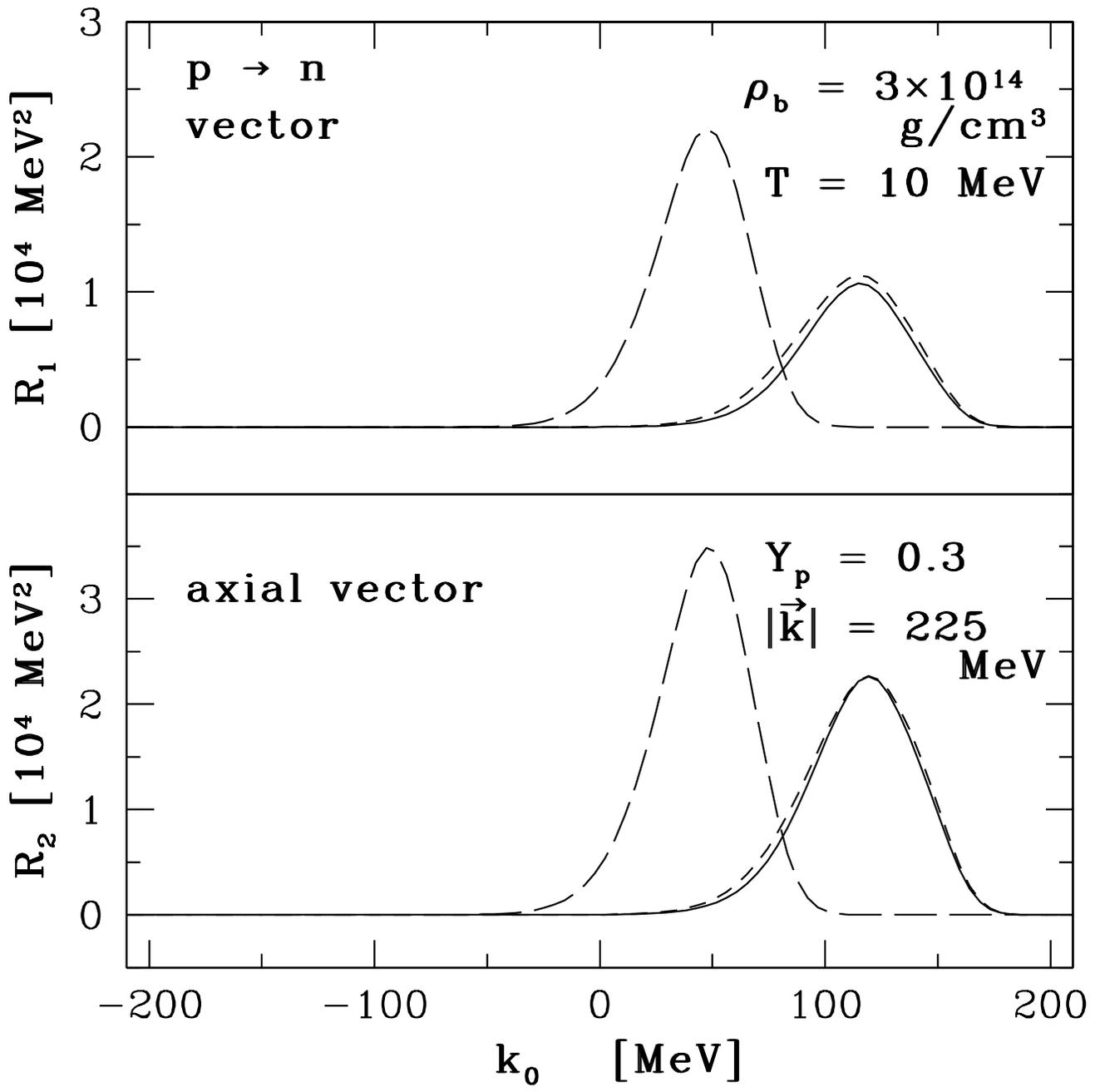
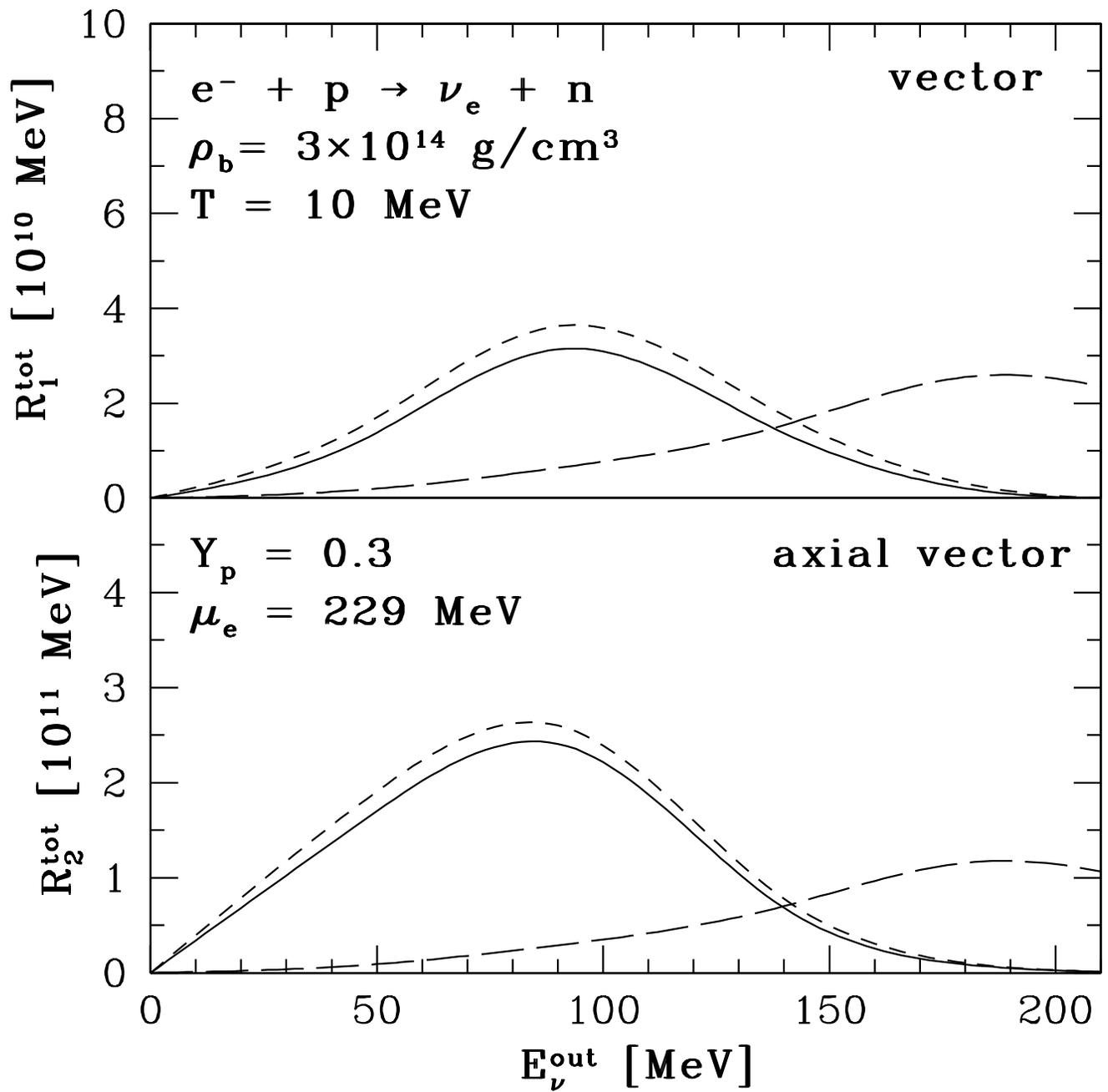
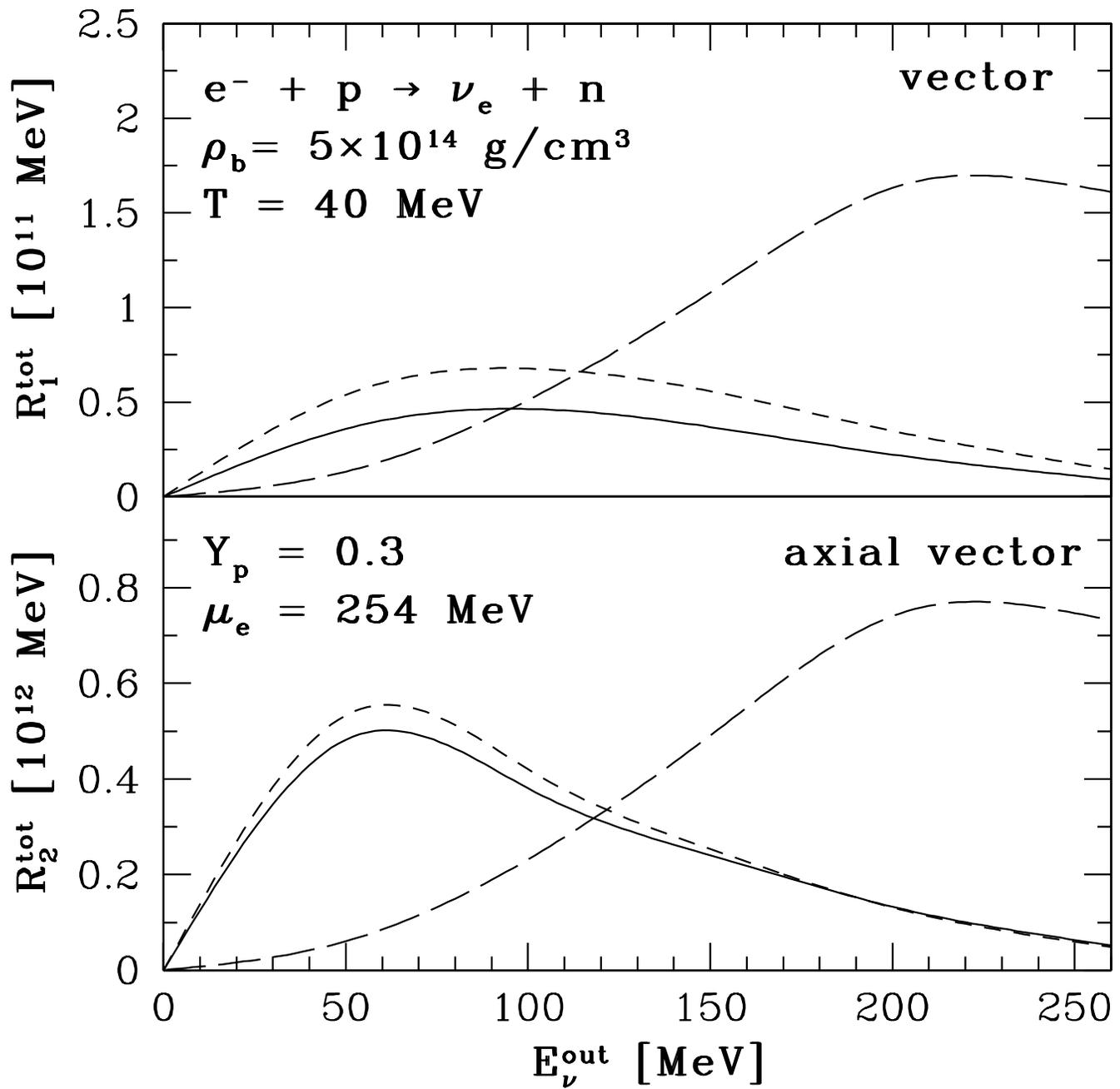
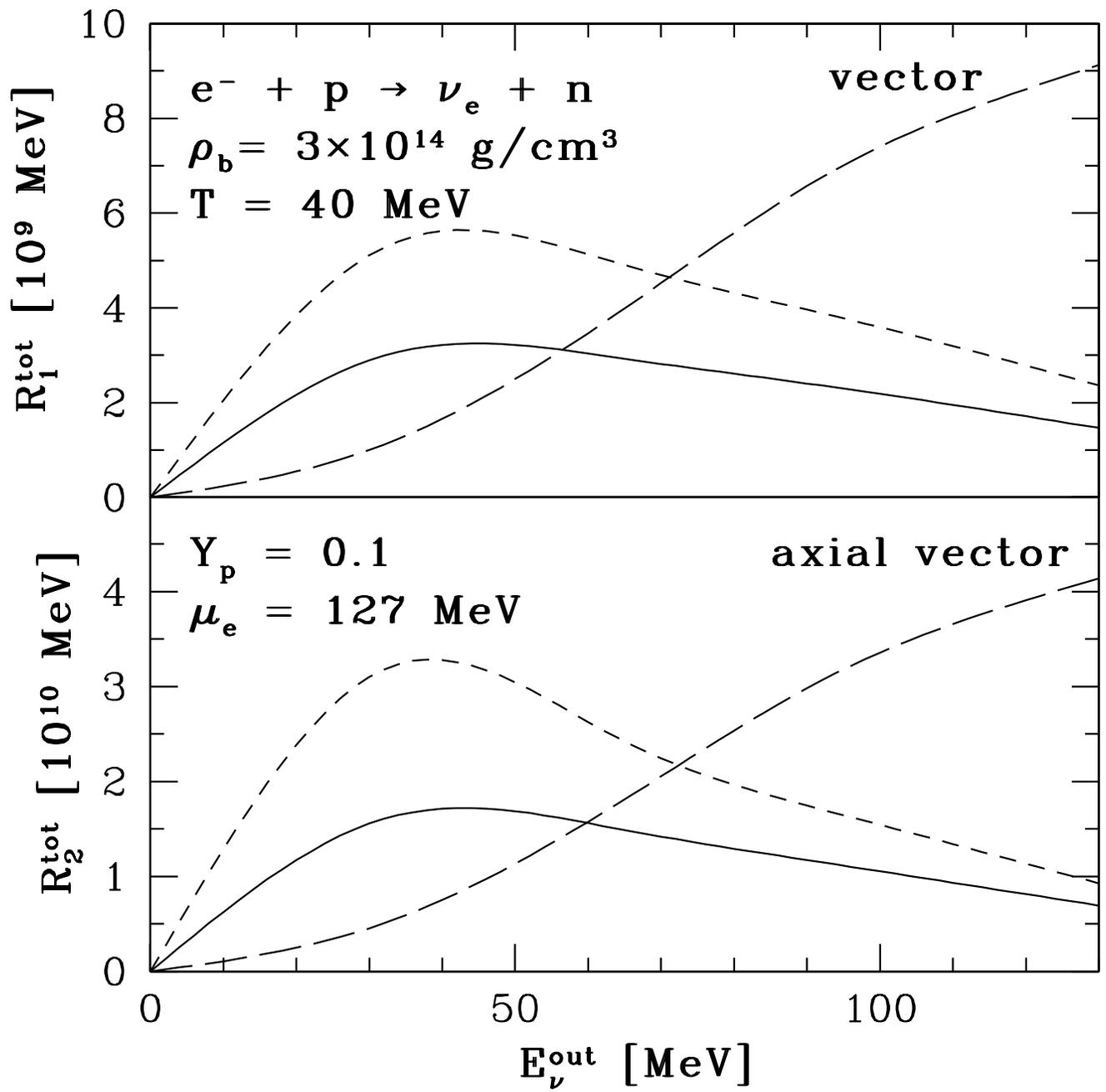
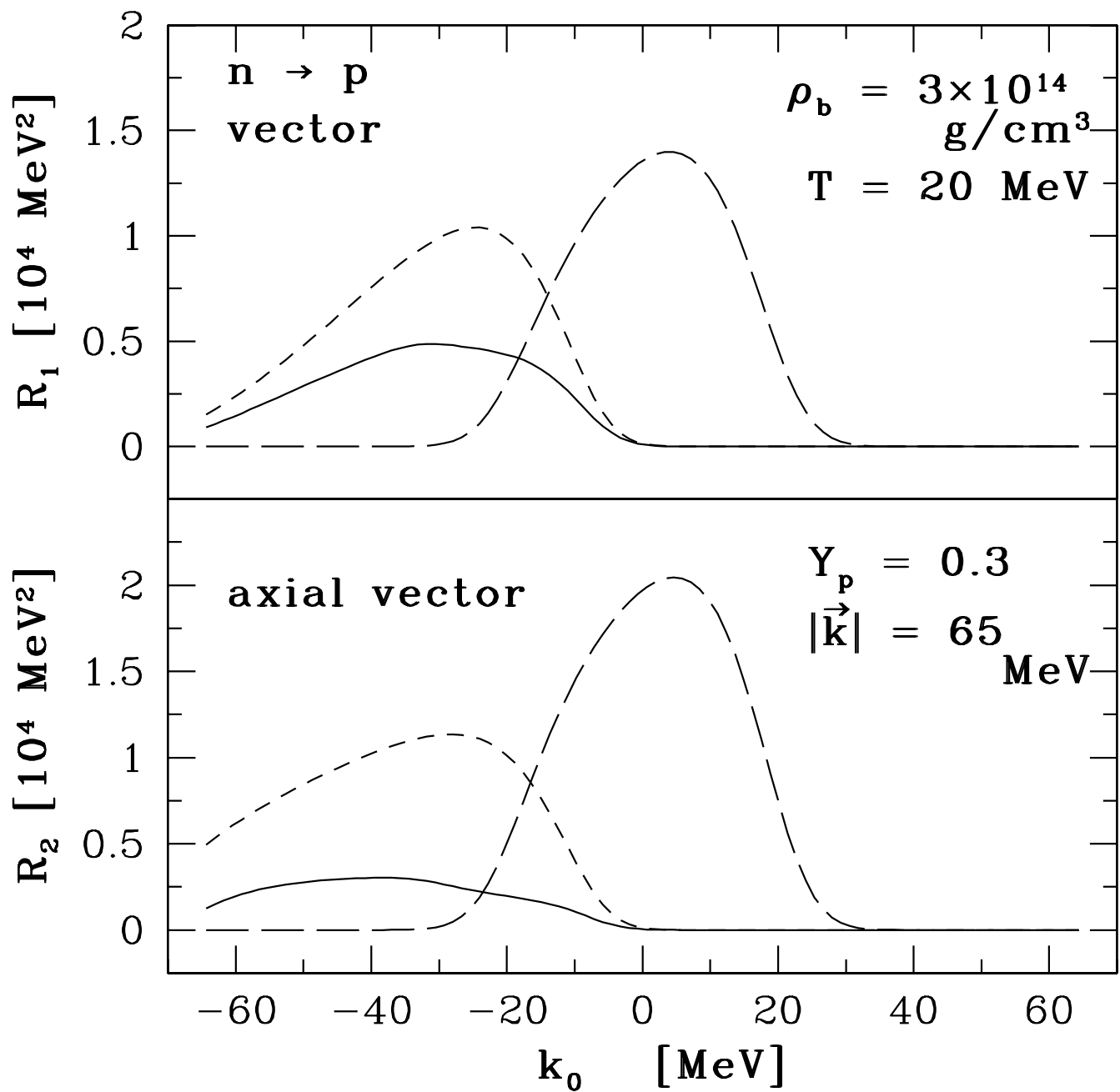
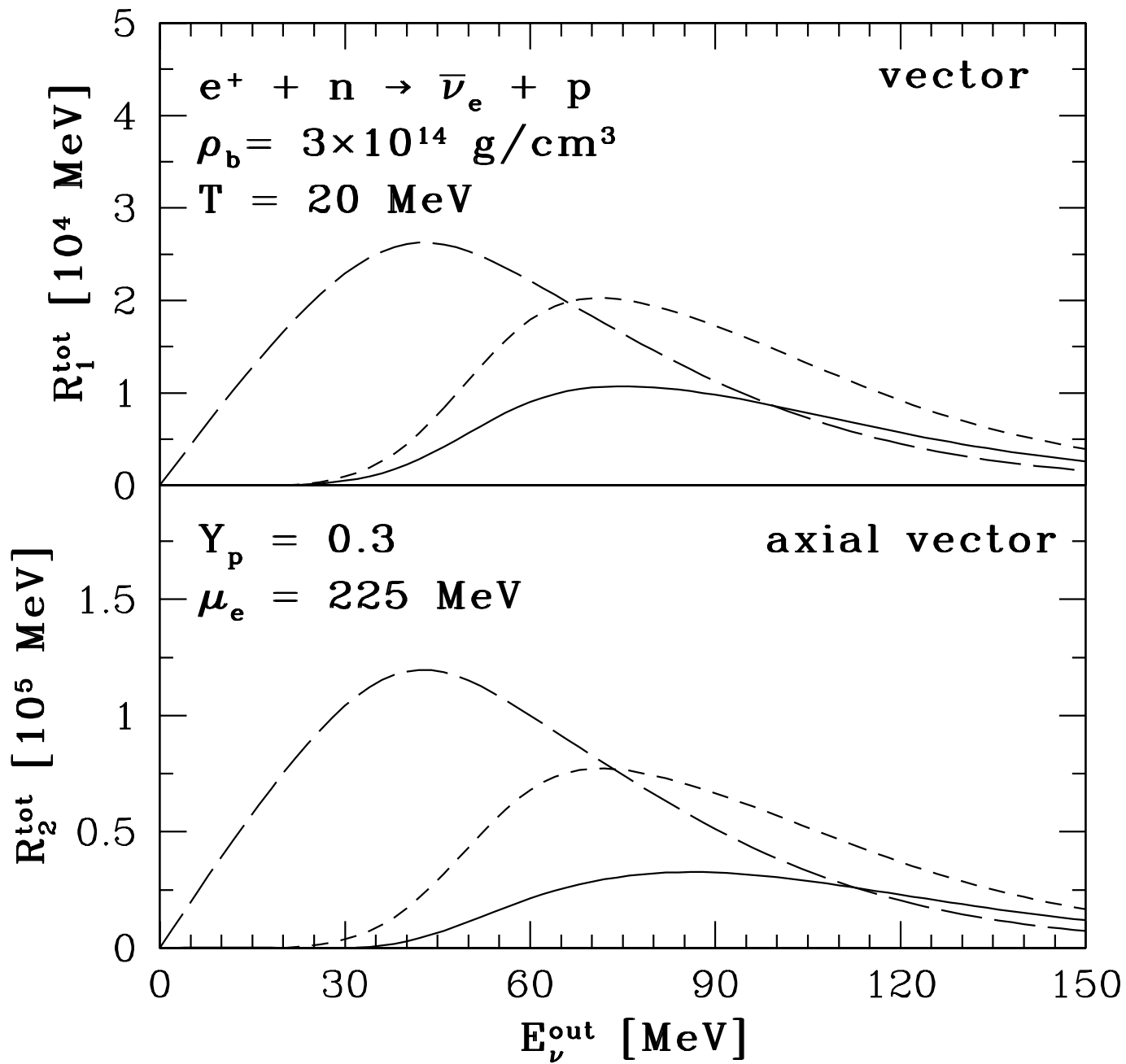
\end{document}